\documentclass[floatfix,twocolumn]{aastex631}

\usepackage{amsmath}
\usepackage{textcomp}
\usepackage{gensymb}
\usepackage{stackengine}
\usepackage{subfigure}

\newcommand{\rvar}{$R_\mathrm{var}$}
\newcommand{\starspot}{\texttt{starspot}}
\newcommand{\celerite}{\texttt{celerite2}}
\newcommand{\lightkurve}{\texttt{lightkurve}}
\newcommand{\stella}{\texttt{stella}}
\newcommand{\astropy}{\texttt{astropy}}
\newcommand{\ernlib}{\texttt{ernlib}}
\newcommand{\IRAF}{\texttt{IRAF}}

\newcommand{\readmultispec}{\texttt{readmultispec}}
\newcommand{\pymc}{\texttt{PyMC3}}

\newcommand{\lhalbol}{$L_{\mathrm{H\alpha}}/L_\mathrm{bol}$}
\newcommand{\lhblbol}{$L_{\mathrm{H\beta}}/L_\mathrm{bol}$}
\newcommand{\lhglbol}{$L_{\mathrm{H\gamma}}/L_\mathrm{bol}$}
\newcommand{\lhdlbol}{$L_{\mathrm{H\delta}}/L_\mathrm{bol}$}
\newcommand{\ha}{H$\alpha$}
\newcommand{\hb}{H$\beta$}
\newcommand{\hg}{H$\gamma$}
\newcommand{\hd}{H$\delta$}
\newcommand{\caii}{Ca~\textsc{ii}~H\&K }
\newcommand{\cak}{Ca~\textsc{ii}~K }
\def\asec{\ifmmode^{\prime\prime}\else$^{\prime\prime}$\fi}
\makeatletter
\newcommand\thefontsize[1]{{#1 The current font size is: \f@size pt\par}}
\makeatother

\defcitealias{garcia_soto_contemporaneous_2023}{G23}

\begin{document}

\title{Short-Term Balmer Line Emission Variability in M Dwarfs}

\author[0000-0001-9828-3229]{Aylin Garc{\'i}a Soto}
\affiliation{Department of Physics and Astronomy, Dartmouth College, Hanover NH 03755, USA}

\author[0000-0002-7119-2543]{Girish M. Duvvuri}
\affiliation{Department of Physics and Astronomy, Vanderbilt University, Nashville TN 37235, USA}

\author[0000-0003-4150-841X]{Elisabeth R. Newton}
\affiliation{Department of Physics and Astronomy, Dartmouth College, Hanover NH 03755, USA}

\author[0000-0002-0583-0949]{Ward S. Howard}
\affiliation{Department of Astrophysical and Planetary Sciences, University of Colorado, Boulder CO 80309, USA}

\author[0000-0002-8047-1982]{Alejandro N{\'u}{\~n}ez}
\affiliation{Department of Astronomy, Columbia University, New York NY 10027, USA}

\author[0000-0001-7371-2832]{Stephanie T. Douglas}
\affiliation{Department of Physics, Lafayette College, Easton PA 18042, USA}

\correspondingauthor{Aylin Garc{\'i}a Soto} 
\email{aylin.garcia.soto.gr@dartmouth.edu}

\begin{abstract}
M Dwarfs make up the majority of stars, offering an avenue for discovering exoplanets due to their smaller sizes. However, their magnetic activity poses challenges for exoplanet detection, characterization, and planetary habitability. Understanding its magnetic activity, including surface starspots and internal dynamos, is crucial for exoplanet research. In this study, we present short-term variability in four Balmer emission lines \ha, \hb, \hg, and \hd\ for a sample of 77 M dwarfs of varying spectral types, and binarity. Stars were observed using the MDM Observatory's Ohio State Multi-Object Spectrograph on the 2.4m  Telescope and the Modular Spectrograph on the 1.3 m Telescope. These data are combined with TESS photometry to explore the connection between spectroscopic and photometric variability. We observe sporadic short-term variability in Balmer lines for some stars, on timescale $\gtrsim$ 15-min, but much shorter than the stellar rotation period. We calculate periods for stars lacking those measurements, re-evaluated the relationship between amplitude (\rvar)-activity relation for the \ha \ line from \citet{garcia_soto_contemporaneous_2023}, and extended our analysis to the \hb, \hg \ and \hd \ lines, which indicates that the relation becomes increasingly dispersed for higher-order Balmer lines. This is consistent with increased intrinsic variability from lower to higher order lines. Additionally, we compute the Balmer decrement, using \hb \ as the fiducial, for stars where we could measure \hg \ and/or \hd. The Balmer decrement can show distinct patterns during white-light flares, with significant differences even for the same star. We also find evidence for dark spots on \object{TIC 283866910}.
\end{abstract}

\section{Introduction} 
\label{sec:intro}
M Dwarfs are the most abundant stars in the galaxy, accounting for 70\% of the stellar population \citep{henry_solar_2006}. Their small sizes and low luminosities, in comparison to the Sun, make them the primary targets in small exoplanet detection efforts such as the Transiting Exoplanet Survey Satellite \citep[TESS;][]{ricker_transiting_2015}. However, their close-in habitable zone means their exoplanets can be subjected to multiple magnetic field-related activities that may obliterate the planets'
 atmospheres such as X-ray irradiation, flares, coronal mass ejections, and filament eruptions \citep[e.g.,][]{khodachenko_coronal_2007,lammer_coronal_2007,segura_effect_2010,muheki_high-resolution_2020}. 

\begin{deluxetable*}{LCCCCCCCCCCCCCC}
\centering
\tablecolumns{14}
\tablecaption{Parameters for some M dwarfs in Our Sample}
\footnotesize
\tablehead{\colhead{TIC} &\colhead{\stackunder{Plx}{(mas)}} & \colhead{\stackunder{K}{(mag)}} & \colhead{\stackunder{$P_\mathrm{rot}$}{(days)}} & \colhead{Ref.}&  \colhead{\stackunder{$R_\star$}{($R_\odot$)}} & \colhead{\stackunder{$M_\star$}{($M_\odot)$}} &\colhead{\stackunder{$T_{\text{eff}}$}{(K)}}  &  \colhead{$R_\mathrm{var}$} & \colhead{In \S \ref{sec:amp}?} & \colhead{In \S \ref{sec:invarb}?} & \colhead{In \S \ref{sec:flares}?} & \colhead{In \S \ref{sec:amber}?}}
\startdata
3664898  & 279.25 & 7.26 & 0.46 & \text{N16} & 0.124 & 0.098 & 2835 & 0.006 & \text{Y} & \text{Y} & \text{N} & \text{N}\\
65673065 & 70.27 & 5.93 & 12.2 & \text{S24} &  0.581 & 0.577 & 3843 & - & \text{N} & \text{Y} & \text{N} & \text{Y} \\
191632569 & 29.56 & 8.24 & 0.72 & \text{T.W.} & 0.509 & 0.511 & 3680 & - & \text{N} & \text{N} & \text{Y} & \text{N} \\
... & ... & ... & ... & ... & ... & ... & ... & ... & ... & ... & ... & ... & ...
\enddata
\label{tab:bigtab}
\tablecomments{
Example references for literature periods: M08 \citep{morin_stable_2008}, N16 - \citep{newton_impact_2016}, S24 - \citep{shan_carmenes_2024},  and ``This Work" or T.W. We include extra columns in the machine-readable version such as 2MASS ID, and the errors for the following columns: Parallax (Plx), K magnitude, $P_\mathrm{rot}$, $R_\star$, $M_\star$, $T_\mathrm{eff}$, and \rvar.}
\end{deluxetable*}

 
\begin{figure}[htp!]
    \centering
    \includegraphics[width=\columnwidth]{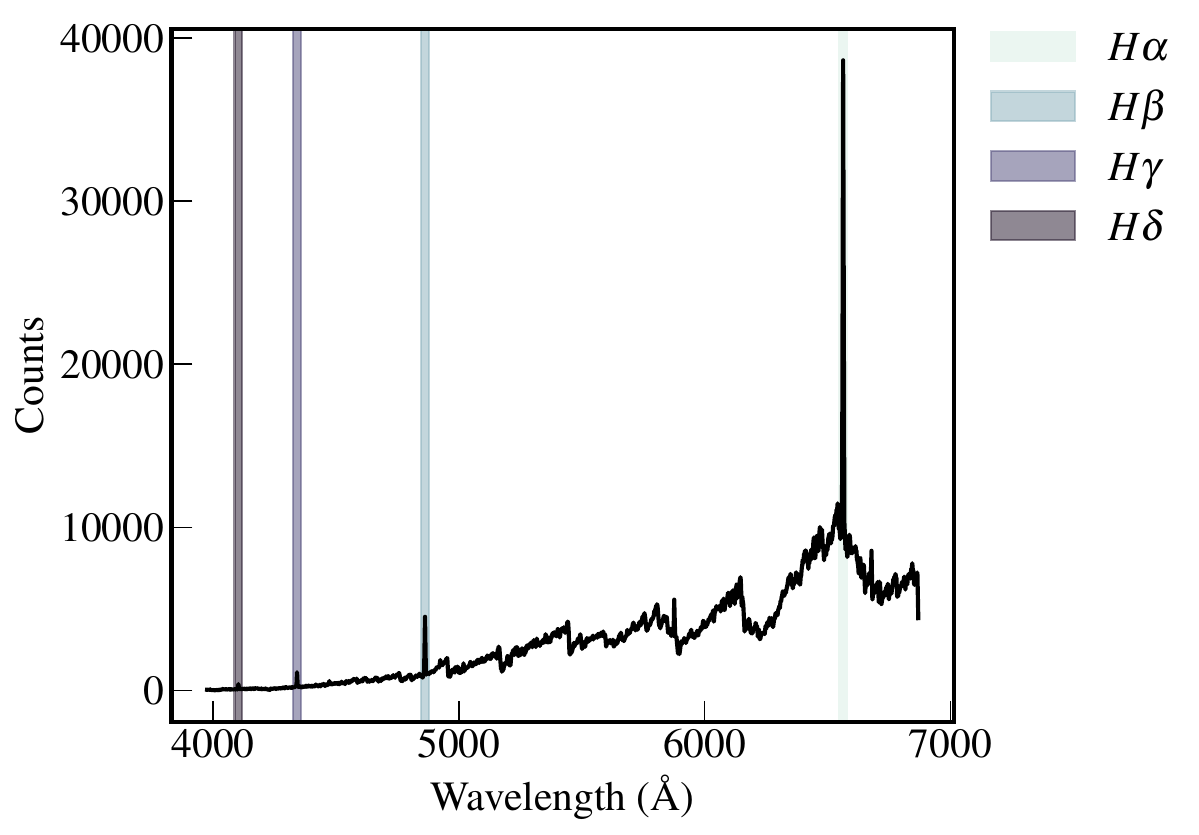}
    \caption{Sample spectrum of a star (\object{TIC 268280825}) with \ha, \hb, \hg \ and \hd, emission lines.}
    \label{fig:samplespec}
\end{figure}

The magnetic dynamo mechanism across low-mass stars is not yet well understood. Observationally, there are no significant changes in the relation between rotation and magnetic activity for the partially convective early M Dwarfs and the later type, fully convective M Dwarfs \citep{wright_stellar_2018,reiners_magnetism_2022}. Coronal emission \citep[e.g.,][]{reiners_evidence_2009,wright_stellar_2018}, chromospheric emission  \citep[\ha, \caii e.g.,][]{vaughan_comparison_1980,mekkaden_rotation_1985, newton_h_2017,boudreaux_ca_2022}, flare activity \citep[e.g.,][]{davenport_kepler_2016,medina_galactic_2022} and magnetic field strength \cite[e.g.,][]{vidotto_stellar_2014,reiners_magnetism_2022} generally show the same relation to stellar rotation in both partially and fully convective M Dwarfs. The relations show two regions: one where activity decreases slightly as stellar rotation slows, and another where activity decreases more significantly with further slowing of rotation \citep[e.g.,][]{mamajek_improved_2008,magaudda_relation_2020,boudreaux_ca_2022}.

Another magnetic field indicator is photometric amplitude in light curves. While the activity related to star spot groups drives chromospheric and coronal emission, their contrast to the stellar photosphere creates troughs in time-series photometric data throughout a stellar rotation period ($P_\mathrm{rot}$). An example of the connection between high-energy emission and starspots is in \citet{notsu_high_2015} and more recently, \citet{yamashita_starspots_2022}. These works show that there is a generally positive relationship between increasing photometric spot amplitude and increasing \caii strength for solar-like stars. A similar relationship is seen for M dwarfs in \ha\ \citep{newton_h_2017,garcia_soto_contemporaneous_2023}.

Even for stars that are always observed to have a baseline level of high-energy emission, emission is enhanced during flares and displays short-term $\approx$15--60 minutes variability \citep[e.g.,][]{gizis_palomarmsu_2002,lee_short-term_2010,kruse_chromospheric_2010,bell_h_2012,medina_variability_2022,duvvuri_fumes_2023,garcia_soto_contemporaneous_2023,kumar_exploring_2023}. Similar behavior is seen in other chromospheric emission lines \citep{duvvuri_fumes_2023,kumar_exploring_2023}. 

In this paper, we build on the dataset of contemporaneous spectroscopic and photometric observations from \citet[][hereafter G23]{garcia_soto_contemporaneous_2023}. We note an error in the $v\sin i$ values as listed in Table 1 in \citet{garcia_soto_contemporaneous_2023}; readers should use the values from the original catalogs \citep{kesseli_magnetic_2018,fouque_spirou_2018,reiners_carmenes_2018} or this work. 
 
 In \S\ref{sec:obs}, we present new observations from 2020 Nov, 2021 Mar, 2021 Oct, and 2021 Dec, which we analyze alongside observations from \citetalias{garcia_soto_contemporaneous_2023} (2019 Oct, 2020 Jan, 2020 Feb, 2020 Dec, and 2021 Jan). In the following sections, we present the instruments used for contemporaneous observations (\S\ref{sec:obs}), the stellar parameters derived for the stars in our sample (\S\ref{sec:params}), and our measurements and analysis, including new values for the Balmer luminosity factor $\chi$, the conversion factor between equivalent width (EW) and line luminosity first introduced by \citet{walkowicz__2004}. In \S\ref{sec:results}, we revisit the amplitude-activity relation (\S\ref{sec:amp}), quantify intrinsic variability in the Balmer lines (\S\ref{sec:invarb}), compare the TESS light curves to the Balmer light curves (\S\ref{sec:flares}), and discuss results found using the RADYN code for flare fitting (\S\ref{sec:radyn_modeling}). Finally, we discuss our findings in \S\ref{sec:conclude}. 

\section{Observations and Data Reduction} \label{sec:obs}
\subsection{Sample}\label{sec:samp}

\begin{deluxetable*}{LCCCCCC}
\centering
\tablecolumns{5}
\tablecaption{Observations of the Sample}
\tablehead{\colhead{Month} & \colhead{Date Ranges (BJD)} & \colhead{Date Ranges (UTC)} &  \colhead{TESS Sectors} & \colhead{Total Objects$^{c}$} & \colhead{Wavelength Range (\AA)}}\
\startdata
\multicolumn{6}{c}{OSMOS}\\
\cline{1-6}
\text{2019 Oct} & 2458762.66475 - 2458767.59502 &	2019/10/06 - 2019/10/11 & 15-17 & 23 & 3975-6869.5\\
\text{2020 Jan} & 2458851.83332 - 2458856.00280 & 2020/01/03 - 2020/01/07 & 19 - 20 & 24^c & 3975-6869.5 \\
\text{2020 Feb} & 2458880.91156 - 2458895.84191&	2020/02/01 - 2020/02/16 & 21 - 22 & 30^c  & 3975-6869.5\\
\text{2020 Dec} & 2459188.73106 - 2459197.90702& 2020/12/05 - 2020/12/14 & 32 & 4 & 3975-6869.5  \\
\text{2021 Jan} & 2459222.86134 - 2459222.89875 & 2021/01/08 & 33 & 2 & 3975-6869.5 \\
\text{2021 Feb} & 2459250.72317 - 2459250.88450 & 2021/02/05 & 34$^{a}$ & 1^c & 3975-6869.5 \\
\text{2021 Mar} & 2459278.76123 - 2459278.89630 & 2021/03/05 & 35$^{b}$& 1^c & 3975-6869.5 \\
\text{2021 Oct} & 2459501.88114 - 2459512.94050 & 2021/10/14 - 2021/10/25 & 43-44 & 9^c & 3975-6869.5\\
\text{2021 Dec} & 2459560.80704 - 2459560.85435 & 2021/12/12 & 46$^{a}$ & 1 & 3975-6869.5 \\
\cline{1-6}
\multicolumn{6}{c}{ModSpec} \\
\cline{1-6}
\text{2020 Nov} & 2459158.50318 - 2459158.50611 & 2020/11/05 & 31-32$^{b}$ & 8^c & 4660-6732  \\
\text{2021 Mar}& 2459278.50528 - 2459283.50252 & 2021/03/05 - 2021/03/10 & 35-36$^{b}$ & 10 &  4660-6732 \\
\text{2021 Oct}& 2459501.50570 - 2459512.50543 &  2021/10/14 - 2021/10/25 & 43-44  & 9^c &  5535-7615 \\
\enddata
\label{tab:obs}
\tablecomments{(a) These runs did not coincide with TESS observations. (b) These runs did not coincide with short-cadence TESS observations but did coincide with 30-minute TESS observations. (c) 29 targets were observed during more than one run. There are 93 unique targets, which are all included in Table~\ref{tab:bigtab}.}
\end{deluxetable*}

As in \citetalias{garcia_soto_contemporaneous_2023}, we chose objects based on targets with $v \sin i$ measurements from \citet{fouque_spirou_2018,reiners_carmenes_2018,kesseli_magnetic_2018} and $P_\mathrm{rot}$ values less than 13 days. As in \citetalias{garcia_soto_contemporaneous_2023}, we aim to overlap our ground observations with TESS observations and limit our targets to only M Dwarfs. However, in this work, we consider spectroscopic data without simultaneous TESS data. 
Furthermore, in \citetalias{garcia_soto_contemporaneous_2023}, we discarded stars with a RUWE $>1.6$ and a TESS contamination ratio (CR) of $<$0.2.\footnote{RUWE is a $\chi$-square metric for fitting models to individual stars, designed to eliminate the correlation between color and magnitude in the Gaia Unit Weight Error meant to flag possible visual binaries. CR is the fractional contribution of neighboring stars relative to the target's flux as per \citet{stassun_tess_2018}.} 
For this paper, we only note these stars as candidate binaries. Thus, our sample contains rapidly rotating stars that are a mix of presumed single stars and likely binaries. Our full dataset is composed of 93 targets and is listed in Table~\ref{tab:bigtab}. We specify which subset of the dataset is used in each section.

77/93 of our targets are M dwarfs (\S\ref{sec:params}) and are analyzed in this work. Because we limit our observations based on $P_\mathrm{rot}$, the majority of our stars are fully convective: this is a natural result of mid-to-late M dwarfs spending more time rapidly rotating and in the saturated regime of the rotation-activity relation.

\subsection{Observations and Spectra Reduction with PyRAF and IRAF}
We utilize the Ohio State Multi-Object Spectrograph (OSMOS; FWHM $R \sim 3.1$ \AA) spectrograph \citep{martini_ohio_2011}, mounted on the 2.4 m Hiltner telescope at MDM Observatory in Arizona, which was also used in \citetalias{garcia_soto_contemporaneous_2023}.  We continue to use the blue VPH grism with an inner 1.2$\asec$ slit and a 4x1k region of interest, resulting in a wavelength coverage of 3975--6869.5 \AA. 

We also use data from the Modular Spectrograph (ModSpec; FWHM $\sim 3.5$ \AA) mounted on the 1.3 m McGraw-Hill at MDM Observatory. We use a grating tilt of about 26.06 degrees and a collimator focus near 670. The data is mostly centered around \ha, thus our wavelength ranges do not include the other Balmer lines of higher order than \ha, except 2020 Nov, and 2021 Mar which include \hb \ (Table~\ref{tab:obs}).

The reduction process for OSMOS utilizes the \texttt{PyRAF} package \citep{science_software_branch_at_stsci_pyraf_2012} and scripts within \href{https://github.com/jrthorstensen/thorosmos}{\texttt{thorosmos}}\footnote{\url{https://github.com/jrthorstensen/thorosmos}} and is described in \citetalias{garcia_soto_contemporaneous_2023}. 

\begin{deluxetable*}{LCCCCCCCCC}
\centering
\tablecaption{Median Equivalent Widths and Luminosities for H$\alpha$, H$\beta$, H$\gamma$, and H$\delta$}
\tablehead{\colhead{TIC} & \colhead{$EW_{H\alpha,relative}$} & \colhead{$EW_{H\alpha}$ } & \colhead{$EW_{H\beta}$} & \colhead{$EW_{H\gamma}$} & \colhead{$EW_{H\delta}$} & \colhead{\lhalbol} & \colhead{\lhblbol} & \colhead{\lhglbol} & \colhead{\lhdlbol}}
\startdata
191632569 &	-4.630 & -4.305 & -6.843 & -10.23 & -8.05	& 1.47 \times10^{-4}  & 1.86\times10^{-4} & 3.62\times10^{-4} & 1.05\times10^{-4}\\
... & ... & ... & ... & ... & ... & ... & ... & ... & ...
\enddata\label{tab:spectra}
\tablecomments{We provide a supplemental machine-readable table for each Balmer line with times, with all EWs, luminosity values, BJD dates, exposure times, and airmass calculated for each object even the discarded ones from \citetalias{garcia_soto_contemporaneous_2023}.}
\end{deluxetable*}

\begin{deluxetable}{LLLLL}
\tablecaption{$\chi$ values for a subset of the Balmer Lines}
\tablehead{\colhead{\stackunder{$T_{\text{eff}}$}{(K)}} & \colhead{\stackunder{$\chi_\mathrm{H\alpha}$}{($\times10^{-5}$)}} & \colhead{\stackunder{$\chi_\mathrm{H\beta}$}{($\times10^{-5}$)}} & \colhead{\stackunder{$\chi_\mathrm{H\gamma}$}{($\times10^{-5}$)}} & \colhead{\stackunder{$\chi_\mathrm{H\delta}$}{($\times10^{-5}$)}}}
\startdata
2300 & 0.673 & 0.076 & 0.034 & 0.039 \\
2400 & 0.735 & 0.048 & 0.052 & 0.067 \\
2500 & 0.547 & 0.052 & 0.068 & 0.107 \\
2600 & 0.726 & 0.086 & 0.110 & 0.158 \\
2700 & 1.032 & 0.152 & 0.176 & 0.223 \\ 
2800 & 1.418 & 0.254 & 0.258 & 0.288 \\ 
2900 & 1.880 & 0.398 & 0.365 & 0.366 \\
3000 & 2.363 & 0.577 & 0.491 & 0.454 \\
3100 & 2.894 & 0.799 & 0.641 & 0.558 \\
3200 & 3.506 & 1.081 & 0.829 & 0.691 \\
3300 & 4.088 & 1.391 & 1.029 & 0.830 \\
3400 & 4.645 & 1.728 & 1.236 & 0.970 \\
3500 & 5.177 & 2.080 & 1.440 & 1.105 \\
3600 & 5.648 & 2.441 & 1.631 & 1.227 \\
3700 & 5.999 & 2.797 & 1.797 & 1.329 \\
3800 & 6.329 & 3.167 & 1.962 & 1.432 \\
3900 & 6.642 & 3.556 & 2.141 & 1.550 \\
4000 & 6.938 & 3.975 & 2.329 & 1.678 \\
4100 & 7.256 & 4.449 & 2.550 & 1.834 \\
4200 & 7.750 & 5.078 & 2.861 & 2.061 
\enddata
\label{tab:chi}
\tablecomments{Using the method from \citet{nunez_factory_2024} and \citet{douglas_factory_2014}. The errors are assigned as 10\% consistent with both of those papers. Note, however, that this paper had different regions as mentioned in Table~\ref{tab:regs}.}
\end{deluxetable}

For ModSpec we manually reduce the data using \texttt{IRAF} \citep{tody_iraf_1986,tody_iraf_1993}. Note that neither OSMOS nor ModSpec requires us to take bias or dark frames because the instrument produces several columns of an overscan region that we use as bias and remove from the frames. Next, we median combine both sky flats and Multi-Instrument System flat or an Internal flat and remove them from the frames.

We fit an aperture around the signal and the sky background, and then a trace, a polynomial, to the center of the aperture. We optimally extract the spectra \citep{horne_optimal_1986}. After this, we create a wavelength solution by marking several lines known within the mercury, xenon, and neon comparison lamp frames, iterating until we achieve a root-mean-square residual of 0.1 \AA. We then dispersion-correct the spectra (Figure~\ref{fig:samplespec}).

Lastly, since the time recorded for spectral observations is the beginning of the observations, we add half of the exposure time to each timestamp for both ModSpec and OSMOS observations to match TESS timestamps, which are recorded at the midpoint of the observations. This process was done before calculating the barycentric julian date with \astropy \citep{astropy_collaboration_astropy_2013,astropy_collaboration_astropy_2018,astropy_collaboration_astropy_2022}.

\section{Stellar Parameters}\label{sec:params}

For stellar radii, we use an empirical relationship between absolute K magnitude ($M_{Ks}$) and radius, expressed as $R_\star = 1.9515 - 0.3520M_K + 0.01680M_K^2$ \citep{mann_how_2015}. Apparent {\it JHKs} magnitudes for all 93 stars are extracted from the 2MASS All-Sky Catalog \citep{cutri_vizier_2003} with photometric quality of AAA, while most of the parallaxes are extracted from {\it Gaia} DR3 \citep{, lindegren_gaia_2021, gaia_collaboration_gaia_2022, gaia_collaboration_gaia_2023}. For three stars (\object[TIC 348385129]{TIC 348385129}, \object[TIC 406670485]{406670485}, and \object[TIC 274127410]{274127410}) the parallaxes are from {\it Gaia} DR2 \citep{gaia_collaboration_gaia_2016, gaia_collaboration_gaia_2018, lindegren_gaia_2018} and for two stars (\object[TIC 252470185]{TIC 252470185} and \object[TIC 428740087]{428740087}) the parallaxes are extracted from \citet{dittmann_trigonometric_2014}. These stars we consider candidate binaries. One star, \object{TIC 268281426}, did not have a parallax and thus is removed from \S~\ref{sec:results}\footnote{Many parameters depend on the parallax, such as radius and mass--the latter consequently needed to calculate \ha\ luminosity. In Table~\ref{tab:bigtab}, those columns are replaced with NaN values.}. Parallax errors, typically in the range of 3--4\%, dominate the final errors on the stellar radii.

For stellar masses $M_\star$, we use the relationship established by \citet{mann_how_2019} for M dwarfs within the range $0.075 < M/M_\odot < 0.70$. Specifically, we use the \href{https://github.com/awmann/M_-M_K-}{\texttt{M\_-M\_K-}}\footnote{\url{https://github.com/awmann/M_-M_K-}} \ Python package to calculate the posterior probability distributions for the masses of our sample and adopt the median and the 68\% confidence interval as our mass and error, respectively.

\begin{deluxetable*}{LCCC}
\centering
\tablecaption{Wavelength Regions for Balmer Emission Lines Analyzed in this Work}
\tablehead{\colhead{Line} &  \colhead{Blue Continuum Region (\AA)} & \colhead{Line Width (\AA)} & \colhead{Red Continuum Region (\AA)}} \
\startdata
H\alpha & 6500 - 6550 &	8-18 & 6575-6626\\
H\beta & 4840-4850 & 8-16  & 4875-4885 \\
H\gamma &4310-4330 & 8-14  & 4350-4370\\
H\delta & 4060-4080 & 6-12  & 4120-4140 \\
\enddata
\label{tab:regs}
\tablecomments{We reference \citet{newton_h_2017} (\ha) and $H\beta, H\gamma, H\delta$ \citep{walkowicz_tracers_2009} and \citet{duvvuri_fumes_2023} to set the continua regions. Line widths are selected visually for each star, some to account for the wings on some emission lines. }
\end{deluxetable*}

While \citet{mann_how_2015} has an absolute G magnitude-$T_\mathrm{eff}$ relation, it is dependent on metallicity \citep{delfosse_accurate_2000, mann_how_2015}, so our effective temperatures are derived using the $T_\mathrm{eff}$--$M_{Ks}$ relation from \citet{pecaut_intrinsic_2013} (Version 2022.04.16). 

Lastly, we exclude 16 stars from the analysis in \S~\ref{sec:results}, for stars with both $T_\mathrm{eff} \geq$ 4000 K and $M_\star$ $\geq$ 0.6 $M_\odot$. The largest $M_\star$ ($\approx$0.42 $M_\odot$) and largest $T_\mathrm{eff}$ ($\approx$3480 K) are for the same star, \object{TIC 307913606}. Thus, there are 77 M dwarfs in our sample. However, the parameters of all 93 stars are included in the machine-readable version of Table~\ref{tab:bigtab}. 

\section{Photometric and Spectroscopic Analysis}\label{sec:methods}
\subsection{TESS photometric lightcurves}\label{sec:tesslc}

We download 20-second or 2-minute cadence data from TESS using the \lightkurve \ package \citep{lightkurve_collaboration_lightkurve_2018, geert_barentsen_keplergolightkurve_2020}. These data are released by the Science Processing Operations Center via the Mikulski Archive for Space Telescopes (MAST) archive--The 20-second data is available at MAST: \dataset[https://doi.org/10.17909/t9-st5g-3177]{https://doi.org/10.17909/t9-st5g-3177}, and 2-minute: \dataset[https://doi.org/10.17909/t9-nmc8-f686]{https://doi.org/10.17909/t9-nmc8-f686} 

We identify flares using \href{https://github.com/afeinstein20/stella}{\stella}\footnote{\url{https://github.com/afeinstein20/stella}}, a convolutional neural-network based flare-detection algorithm \citep{feinstein_flare_2020}. Using models trained on 2-min TESS data, \stella \ scores the shapes in a light curve between 0 (non-flare) and 1 (flare). Although the scoring output from \stella\ is not a direct probability, it provides a rough but acceptable estimator for flare detection. Instead of using a single model, we choose an ensemble of models (each with a different initial seed) and pick a threshold of 0.5 to represent a flare. 

\subsection{Stellar rotation}

We first remove the flares we identified with {\stella}. We then bin the BJTD to 10 minutes to mitigate smaller flares that might have been missed by \stella, and to improve computational efficiency.  Lastly, we run the binned data through our version of \starspot \citep{angus_ruthangusstarspot_2021,angus_agarciasoto18starrotate_2023}\footnote{\url{https://github.com/agarciasoto18/starrotate}}. \starspot \ is a package used to measure rotation periods and other parameters, such as amplitude, from photometric light curves using Gaussian Processes. As described in \citetalias{garcia_soto_contemporaneous_2023}, the Gaussian process kernel function includes two stochastically-driven simple harmonic oscillators representing the primary and secondary modes of the period. Additional terms account for variability amplitudes and damping. We fit for jitter and/or long-term trends in the light curves. For the stars with the fiducial period from the literature (listed in Table~\ref{tab:bigtab}), a truncated normal prior with a width of 30\% is placed on $\log{P_\mathrm{rot}}$ to prevent unnecessary parameter space exploration near $P_\mathrm{rot}/2$. For the stars without a literature period, we run \lightkurve's Lomb-Scargle function to apply an initial period. We perform a maximum a posteriori (MAP) fit using \starspot\ and then calculate the residuals. Data is retained when the absolute value of the residual is smaller than four times the residual's root mean square (RMS). Finally, using the masked data, we sample the posterior of the stellar rotation model using \pymc\ as implemented in \starspot.

\begin{figure*}[htp!]
    \centering
    \includegraphics[width=.85\columnwidth]{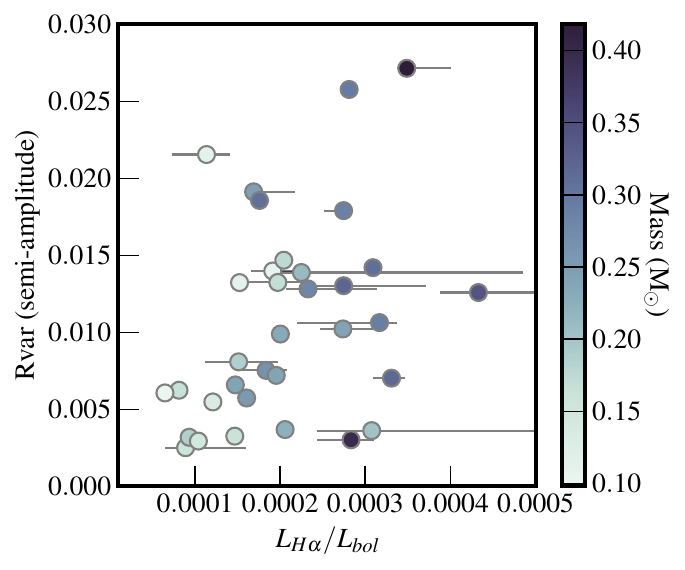}
    \includegraphics[width=.85\columnwidth]{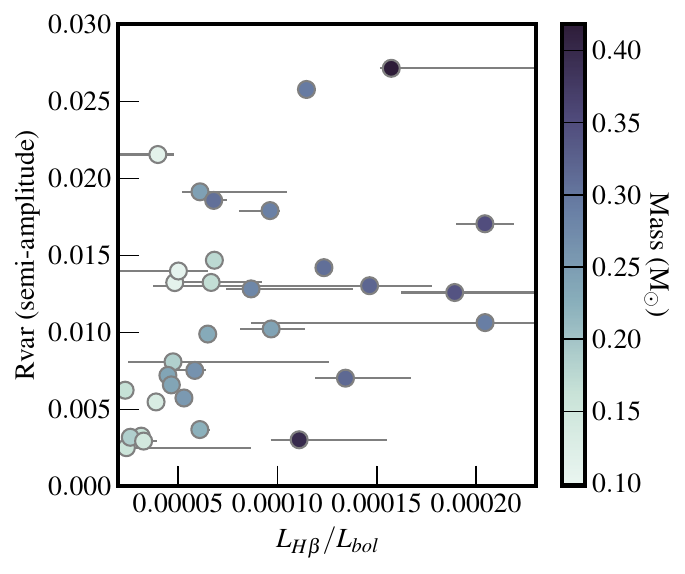}
    \includegraphics[width=.85\columnwidth]{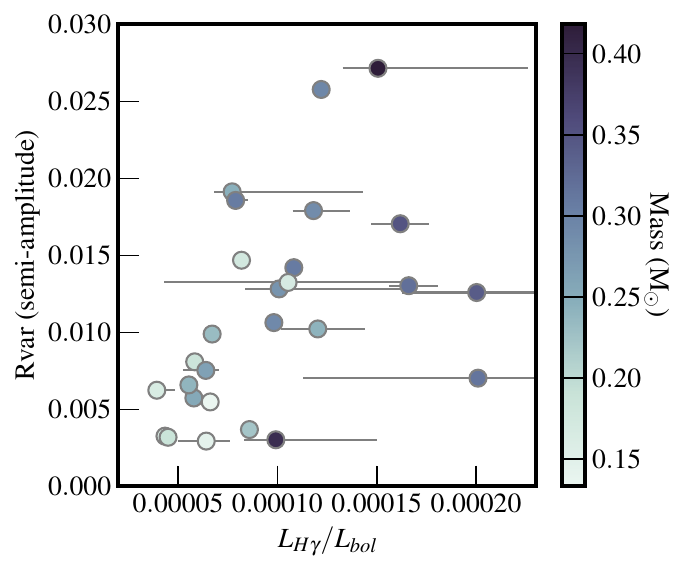}
    \includegraphics[width=.85\columnwidth]{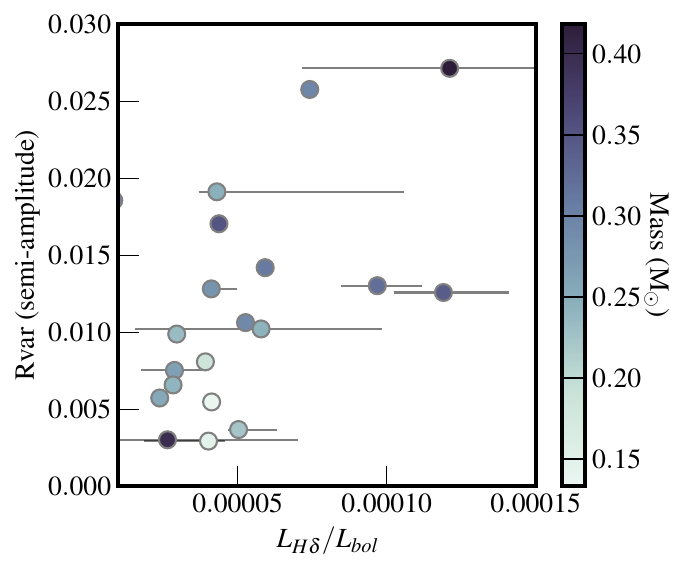}
    \caption{\rvar \ versus the median \lhalbol, \lhblbol,\lhglbol \ and \lhdlbol. We applied the RUWE and CR criteria along with a $M_\star$ cut of 0.45 $M_\odot$. To show the variability of the Balmer lines, instead of the error bars plotted for Balmer luminosity, we plot the range of luminosities (min to max).}
    \label{fig:rvar}
\end{figure*}

\subsubsection{\texorpdfstring{\rvar}{Rvar} }\label{sec:rvar}

We calculate the semi-amplitude of the TESS light curves, after removing the flares, using the \rvar \ method \citep{basri_photometric_2010, basri_photometric_2011,mcquillan_statistics_2012, mcquillan_measuring_2013}: subtracting the 5th from the 95th percentile of the flux, and dividing by two. To determine the statistical uncertainty of \rvar, we vary the flux assuming a Gaussian distribution centered at the measured flux with standard deviation equal to the reported error and record the sampled flux's \rvar. We then calculate the difference between the median and 16\textsuperscript{th}/84\textsuperscript{th} percentiles of the sampled \rvar \ measurements and report the higher of the two differences as the \rvar \ uncertainty. We note that systematic uncertainties likely dominate the error budget, but we do not account for these in this paper.

\subsection{Spectroscopy and Balmer line emission}

\subsubsection{Equivalent widths for Balmer lines}

We trim our spectra to ranges containing the Balmer-series lines \ha, \hb, \hg, and \hd\ for our calculations of equivalent width (EW). The standard definition of EW is applied ($EW_{\text{measured}}$; equation 2 in \citetalias{garcia_soto_contemporaneous_2023}), summing partial pixels assuming uniform illumination. We report emission as negative EW. 

For \ha\ only, we account for the basal absorption contribution of the \ha \ line ($EW_\text{basal}$), derived as a function of $M_\star$ in \citet{newton_h_2017}, by subtracting it from the measured value to obtain the final relative value ($EW_{\text{relative}} = EW_{\text{measured}} - EW_{\text{basal}}$). This relative EW will be referred to as $EW_\mathrm{H\alpha}$ from here on out and is used for subsequent calculations, including \lhalbol. 

For the continuum regions, we utilize the work of \citet{newton_h_2017} and \citet{walkowicz_tracers_2009} as well as \citet{duvvuri_fumes_2023}. Line widths for each star are visually selected to allow for varying wing sizes on some emission lines.

\subsubsection{New \texorpdfstring{$\chi$}{x} factor for \texorpdfstring{$L_{n}/L_{bol}$}{Ln/Lbol}}

We calculate the luminosities of each emission line relative to the bolometric luminosity, \texorpdfstring{$L_{n}/L_\mathrm{bol}$}{Ln/Lbol}, where $n$ represents the different Balmer lines. We obtain non-flux calibrated spectra through the $\chi$ factor \citep{walkowicz__2004}. The authors introduce a factor $\chi$, where $\chi = f_0/f_\mathrm{bol}$. Here, $f_0$ denotes the flux of the continuum adjacent to the line of interest (see Table~\ref{tab:regs}). $f_\mathrm{bol}$ is the bolometric flux derived from the apparent bolometric magnitude and the bolometric correction (BC). Therefore, we can compute the luminosity by

\begin{equation}
L_{n}/L_\mathrm{bol} = \chi_{n} \times - EW_{n}    
\end{equation}

where $n$ indicates \ha, \hb,  \hg, and \hd, and EW are in units of Angstroms.  
We follow the method from \citet{douglas_factory_2014} and \citet{nunez_factory_2024}, which used templates from PHOENIX ACES stellar models \citep{husser_new_2013} to derive  $\chi$.

Our $\chi$ values are listed in Table~\ref{tab:chi} for \ha, \hb, \hg, and \hd \ in terms of $T_\mathrm{eff}$, and have estimated errors of 10\%. Compared to \citet{nunez_factory_2024}, the largest difference is for stars with $T_\mathrm{eff}$ $<$2900 K (10--20\%); the difference is much lower, $\lesssim$2--10\%, for hotter stars. These differences are due to the $\chi$ values' dependence on the defined continua regions (Table~\ref{tab:regs}). 

Our $\chi_\mathrm{H\alpha}$ and $\chi_\mathrm{H\beta}$ values match those in  \citet[][$\chi_{WH08}$]{west__2008} for $T_\mathrm{eff}$ $\lesssim$3100 K. At higher $T_\mathrm{eff}$ our $\chi$ values are smaller than those of WH08, with the difference increasing up to 3800 K, the maximum temperature considered by WH08. The differences range from 25--55\% for $\chi_\mathrm{H\alpha}$, 15--57\% for $\chi_\mathrm{H\beta}$ at 3800 K, and $<$20\% for $\chi_\mathrm{H\gamma}$ at 3700 K. The same trend is noted in \citet{douglas_factory_2014} for their $\chi_\mathrm{H\alpha}$ values when compared to \citet{west_spectroscopic_2004} and \citet{west__2008} and they attribute the differences to an error in calibrating the bolometric luminosity in the previous work.

\subsubsection{Balmer Decrement, the ratio of Balmer lines to \texorpdfstring{H$\beta$}{Hb}}

The Balmer decrement is a way to describe the energy or flux of various chromospheric lines over a fiducial line, usually \hb\ or \hg\ \citep[]{woolley_Balmer_1936}. This metric has been used to investigate the origin of the chromospheric intrinsic variability \citep[e.g.,][]{houdebine_observation_1994,allred_radiative_2006,kowalski_hydrogen_2017,duvvuri_fumes_2023}. In a hot and dense chromosphere, a flare causes higher-order Balmer lines to exhibit larger increases in flux relative to lower-order lines: as the lower-order lines become more optically thick, more energy escapes through the less-optically thick higher-order lines \citep{cram_model_1979, allred_radiative_2006}. This causes the Balmer decrement to become shallower.


We calculate the Balmer decrement relative to \hb\ in this work. 

\section{Results}\label{sec:results}

In this section, the 77/93 stars are considered, of which different subsamples are used in each subsection. The subsamples are labeled in Table \ref{tab:bigtab} and the selection criteria for each subsample are explained within each subsection.

\begin{figure}[htp!]
    \centering
    \includegraphics[width=\columnwidth]{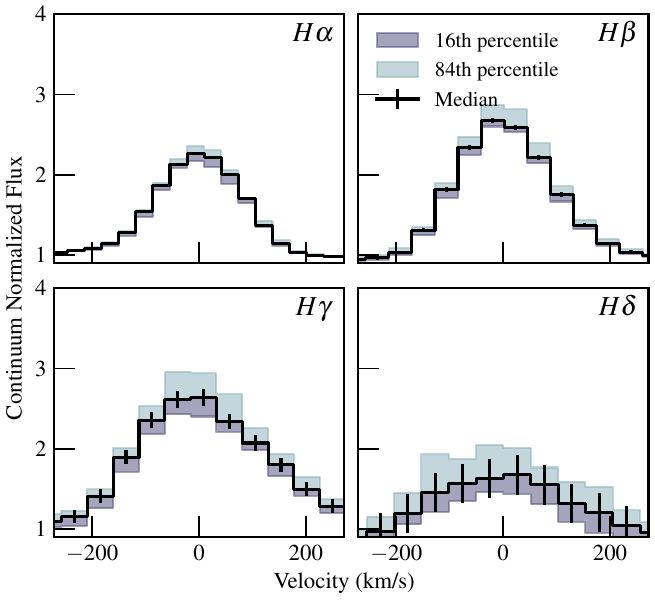}
    \caption{Spectra snippets for \object{TIC 178947176} centered on each Balmer line. The black line represents the median spectrum for each Balmer line, and the shaded regions are the 16th and 84th percentile, indicating the variability of each spectral line. The error bars are the median measurement error of flux values at that wavelength.}
    \label{fig:VarbVel}
\end{figure}

\subsection{No strong amplitude-activity relation in Balmer lines}\label{sec:amp}

We maintain the same criteria as we did in \citetalias{garcia_soto_contemporaneous_2023}, in which we cut the sample to RUWE $\leq$ 1.6, CR $\leq$ 0.2, and $M_\star$ $\leq$ 0.45 $M_{\odot}$ to avoid binaries and influence of nearby stars in the \rvar \ measurement (Figure~\ref{fig:rvar}). This results in 35 stars in our subsample for \ha, 33 for \hb, 26 for \hg, and 20 for \hd. In \citetalias{garcia_soto_contemporaneous_2023}, there were 30 stars for \ha\ and we did not consider the other lines.

First, we repeat the test from \citetalias{garcia_soto_contemporaneous_2023} and calculate Spearman's rank correlation coefficient ($\rho$) and its p-value to determine whether the amplitude is correlated with activity. We incorporate the errors in both the amplitude array and the activity array. As in \citetalias{garcia_soto_contemporaneous_2023}, we find that the correlation is weak and positive, where $\rho =0.394_{-0.169}^{+0.150}$. However, the p-value is large (p = $0.019_{-0.019}^{+0.176}$ or p $ < 0.176$), which means that there is no statistically significant relation between the two parameters.

Similar to \ha, there is a positive, albeit weak, Spearman's rank correlation between \hb\ luminosity and \rvar, with $\rho = 0.429_{-0.162}^{+0.145}$, and between \hg\ luminosity and \rvar, with $\rho = 0.490_{-0.179}^{+ 0.156}$. In both cases, there is no significant evidence against the null hypothesis, with p-values of p $ = 0.012_{-0.012}^{+0.120}$ (i.e., p $ < 0.120$) and p$ = 0.011_{-0.011}^{+0.111}$ (i.e., p $ < 0.111$), respectively.

In contrast, there is no relationship observed between \hd\ luminosity and \rvar, with $\rho = 0.359_{-0.247}^{+0.215}$ and p $ = 0.118_{-0.110}^{+0.448}$ (i.e., p $ < 0.448$). However, as the analysis on \hd \ only includes 20 stars, 
small number statistics and intrinsic variability (as discussed in \S~\ref{sec:invarb}) is likely to contribute to the different result compared to the other Balmer lines. 

\begin{figure*}
    \centering
    \includegraphics[width=\columnwidth]{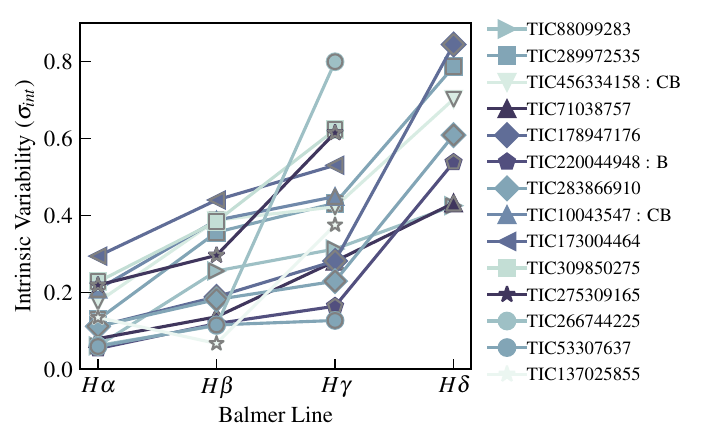}
    \includegraphics[width=\columnwidth]{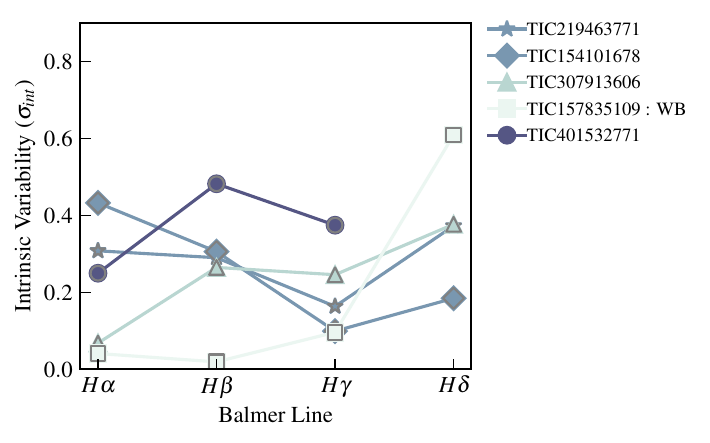}
    \caption{Intrinsic variability of a subsample of stars with more than six spectra, as calculated by Equation~\ref{eqn:invarb}. The colors represent the spectral type (M3-M5) and the shape of the markers represents different objects. An alternative visualization of the same data can be seen in Figure~\ref{fig:hist}. The following labels indicate binarity type: B - Binary, WB - Wide Binary, CB - Candidate Binary or RUWE $>$ 1.6}
    \label{fig:InVarb}
\end{figure*}

\begin{figure}[htp!]
    \centering
    \includegraphics[width=.8\columnwidth]{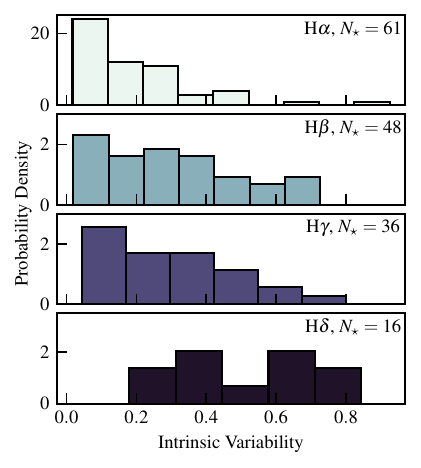}
    \caption{Histogram showcasing the density of the sample plotted against the intrinsic variability in Figure~\ref{fig:InVarb}. We could visibly see a shift in the maximum value of the intrinsic variability towards the high-order Balmer lines.}
    \label{fig:hist}
\end{figure}

\subsection{Larger Intrinsic Variability In Higher Order Balmer Lines}\label{sec:invarb}

In this section, we select stars with more than six spectral observations. The blue side of the spectra sometimes suffered from fringing. We exclude measurements of the bluer Balmer lines (\hg \ or \hd) for stars showing fringing (an example of systematic issues) near or on those Balmer lines.

In Figure~\ref{fig:VarbVel}, spectra centered on each Balmer line with shaded regions indicate the range in continuum normalized flux for stars with multiple observations. We use \object{TIC 178947176} as the example. These figures demonstrate high variability in the higher-order Balmer lines. However, this also includes the impact of decreased signal-to-noise towards the bluer wavelengths. Thus, to further explore this result we define a fractional intrinsic variability parameter, $\sigma_{\text{intrinsic}}$:
\begin{equation}
\sigma_{\text{intrinsic}} = \frac{\sqrt{\sigma_{\text{observed}}^2 - \sigma_{\text{measurement}}^2}}{|\text{Median}(\text{measurement})|}    
\end{equation}\label{eqn:invarb}
where $\sigma_{\text{measurement}}$ is the error in EW and $\sigma_{\text{observed}}$ is half of the difference between the 16th and 84th percentiles of the measured EWs. Lastly, we divide by the median EW to enable comparison across stars.

We plot $\sigma_{\text{intrinsic}}$ against each Balmer line in Figure~\ref{fig:InVarb} and we showed the distribution of $\sigma_{\text{intrinsic}}$ for each Balmer line in Figure~\ref{fig:hist}. Generally, higher-order Balmer lines are increasingly more intrinsically variable, as seen in \citet{duvvuri_fumes_2023}.

\subsection{Discordance between spectroscopic and photometric flares}\label{sec:flares}


A flare occurs when a magnetic field loop breaks: part of the loop snaps off and releases energy, while the other part reconnects at a lower energy state. Some of the released energy heats the chromosphere, leading to enhanced chromospheric emission lines. This heating also produces bright, elongated structures in the chromosphere known as flare ribbons \citep{cheng_numerical_1983,doschek_numerical_1983}. These ribbons form at the foot points of the field loops and often move apart as the flare progresses, reflecting the ongoing reconnection process at higher altitudes in the corona \citep{kazachenko_invited_2022}. The separation between the ribbons provides valuable information about the magnetic field topology and the energy release dynamics of the flare.

\citet{medina_variability_2022} proposes that the dominant source for \ha \ short-term variability is low-energy flares, some of which go undetected by TESS. This section features three stars with simultaneous TESS 20 s cadence data and spectroscopic data of various cadences. 

We search for white-light flares using the typical profile of a flare, a fast rise followed by a slow decay. We note that some flares identified using \stella \ are below the threshold we adopted for detection (\S\ref{sec:tesslc}), such as in the second half of night one for \object{TIC 415508270}. We then compare the TESS light curve to the spectroscopic light curve. 

\begin{figure*}
    \centering
    \includegraphics[width=.75\columnwidth]{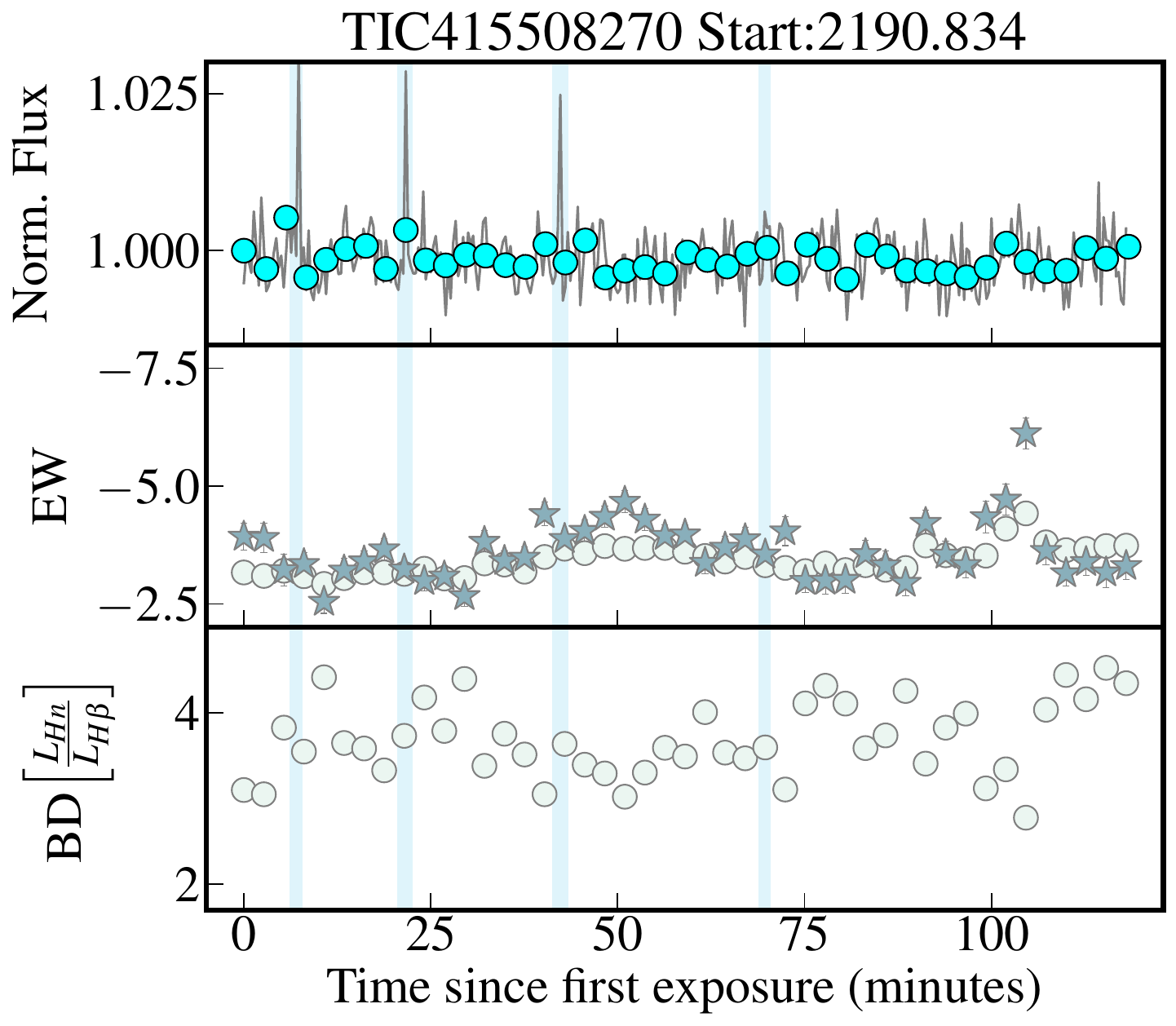}
    \includegraphics[width=.75\columnwidth]{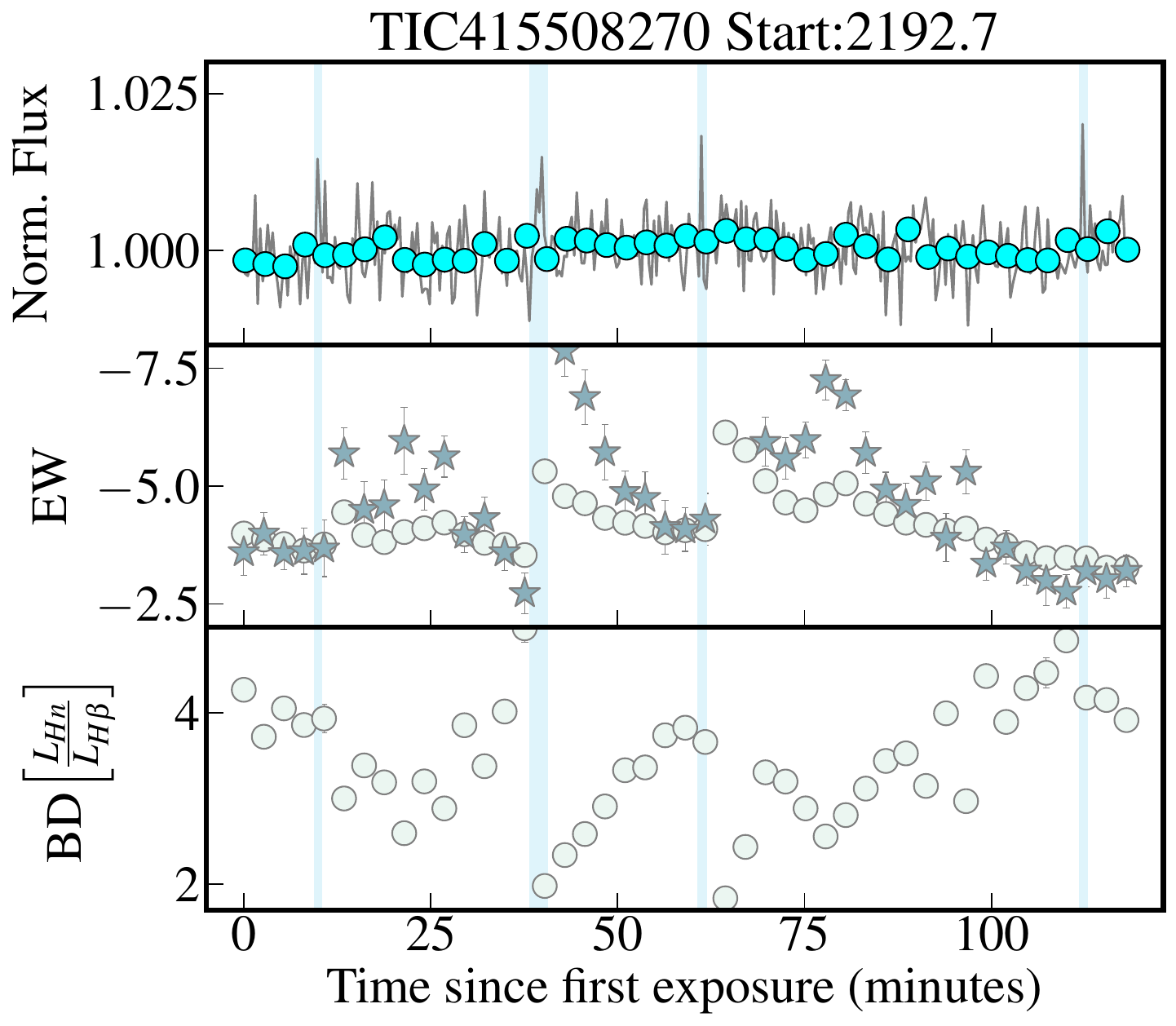}
    \includegraphics[width=\columnwidth]{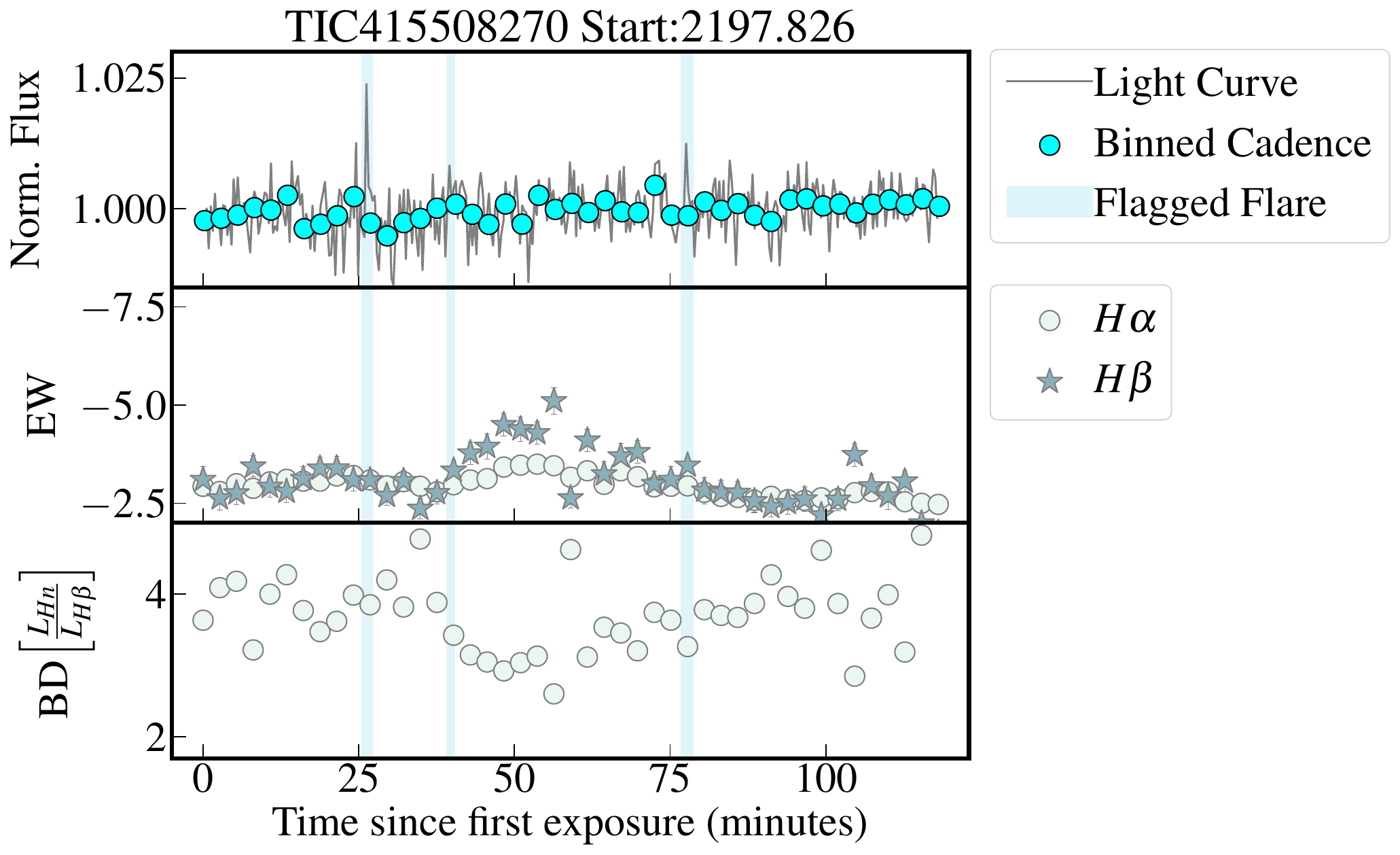}
    \caption{Data from three different observation nights (the three panels) for \object{TIC 415508270}. Within each panel, the top subpanel shows the TESS 20-second cadence light curve plotted against the time since the first spectroscopic observation. We bin the TESS data to match the cadence for the spectroscopic data, which was indicated with cyan circles. We also highlight flares detected by \stella \ with blue-shaded vertical regions. The middle subpanel shows the EW light curve for the detected Balmer lines. Each symbol represents a different Balmer line. The bottom subpanel shows the Balmer decrement measurements, and the luminosity of the Balmer lines over the luminosity of \hb. On the second night (top right panel), the y-axis limit of EW = $-8$ truncates the peak of the flare within the blue-shaded region. We apply this limit to better highlight the subtler variability. Flares are seen in the TESS light curve on all three nights; however, clear flare enhancement signatures in the Balmer lines are only seen on night two.}
    \label{fig:BD415time}
\end{figure*}

\subsubsection{TIC 415508270}

In Figure~\ref{fig:BD415time}, we plot three subplots per observation night, plotted against the minutes since the start of that night's spectroscopic data. The top plot is the TESS short-cadence light curve, in which we bin the data corresponding to the cadence of the spectroscopic data, as represented by the cyan circles. The middle plot represents the equivalent width data for the detected Balmer lines and the bottom plot represents the Balmer decrement, over $L_\mathrm{H\beta}$. The flares found in the TESS data with \stella \ are shaded in blue. 

We show the results for the fully convective M Dwarf, \object{TIC 415508270} in Figure \ref{fig:BD415time}. For this star, we only detect \ha \ to \hb\ \ and so do not look at the higher order lines. The Balmer lines show coherent variability on $\approx$30-minute timescales that are not generally correlated with the TESS light curve.

On the second night, however, we observe clear flares in the Balmer lines. We see a rapid rise and steady decay characterizing the flares. The second flare, around the 35-minute mark, displays a dip in Balmer emission before a complex, long-duration flare event. Pre-flare dips have been observed in the literature for other M dwarfs in a variety of ways. This includes YZ CMi \citep[e.g.,][]{rodono_negative_1979,doyle_rotational_1988}, dMe binary pairs in EQ Peg \citep{giampapa_preflare_1982} and FF And \citep[e.g.,][]{peres_low_1993}, in photometric bands. \citet{leitzinger_search_2014} saw a decrease in the blue wing of their \ha \ profile for \object{2MASS J00032088-3004472}, a potential M dwarf member of Blanco-1. These decreases in radiation could be due, for example, to the destabilization of filaments \citep{giampapa_preflare_1982} or co-rotating prominences \citep[e.g.,][]{collier_cameron_fast_1989}.

\begin{figure*}
    \centering
    \includegraphics[width=.75\columnwidth]{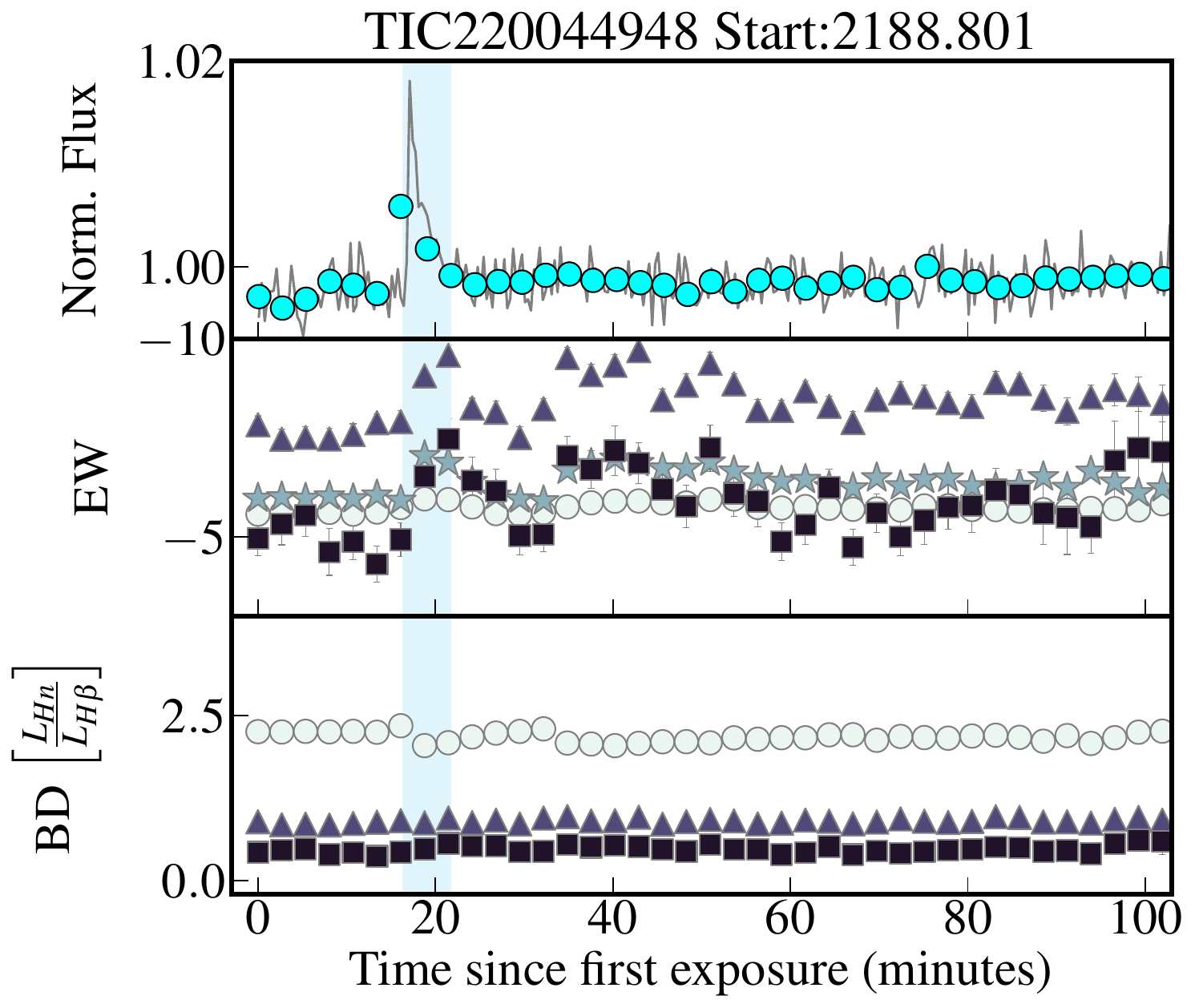}
    \includegraphics[width=.75\columnwidth]{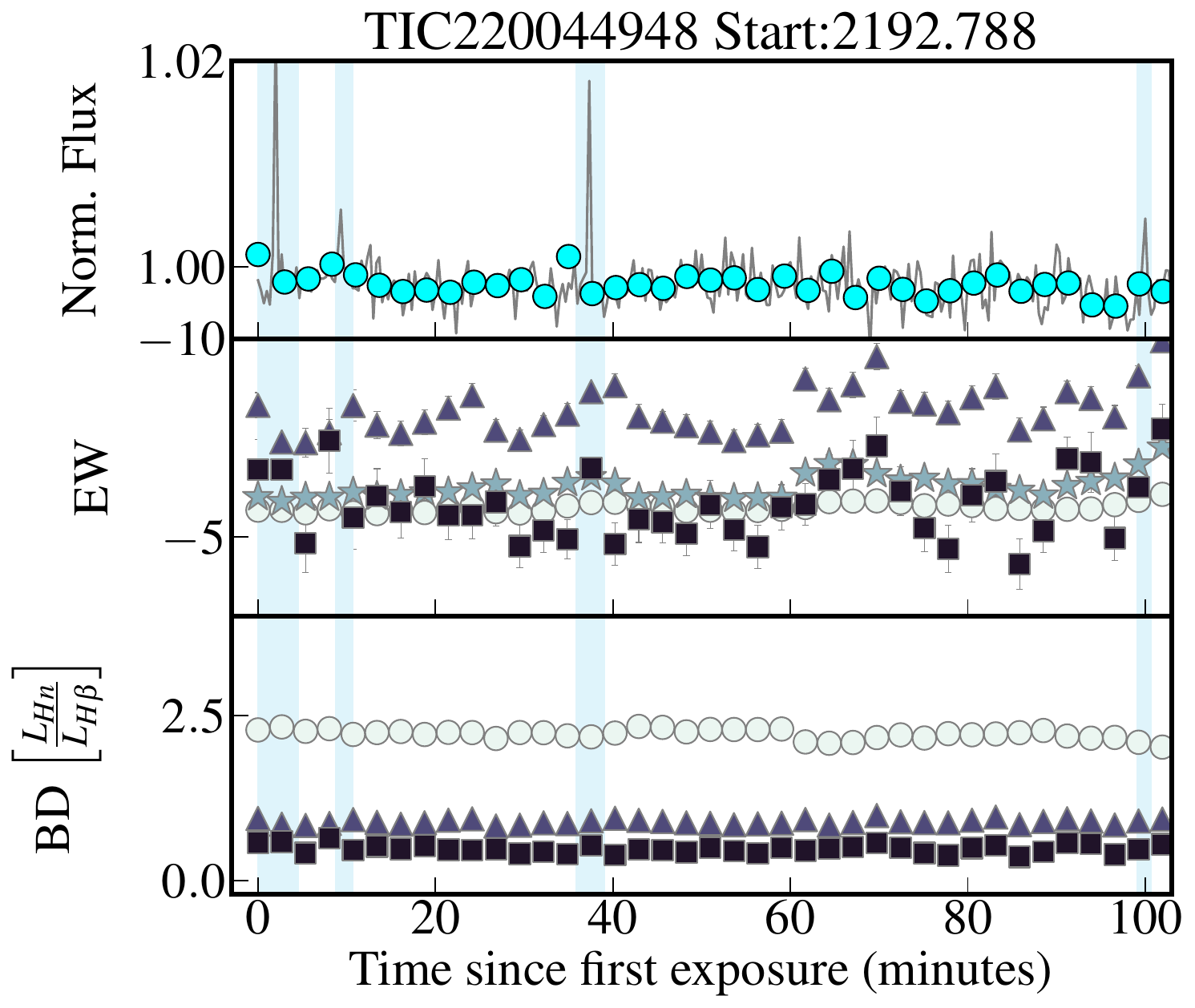}
    \includegraphics[width=\columnwidth]{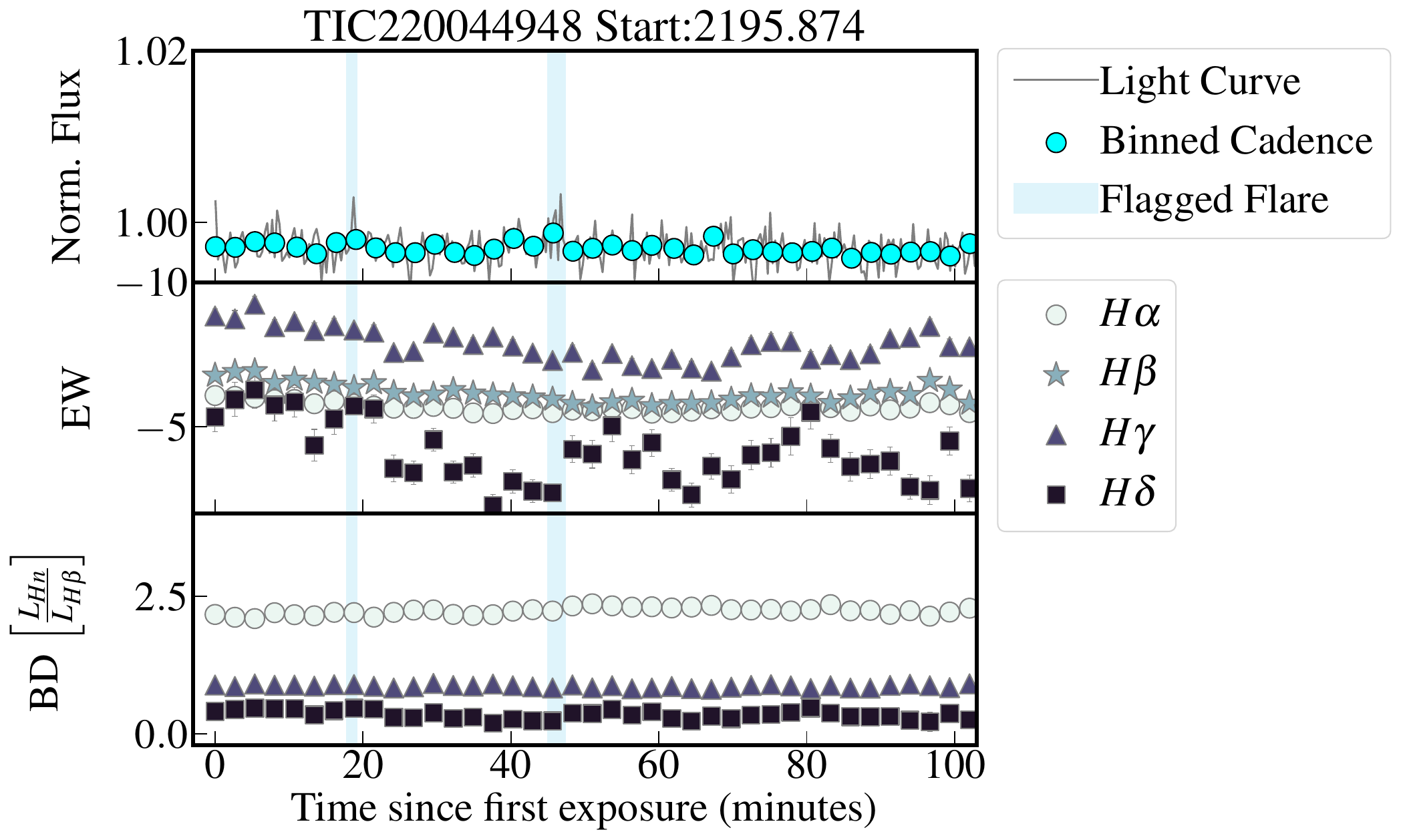}
    \caption{Same as Figure~\ref{fig:BD415time}, but for \object[TIC 220044948]{TIC 220044948}. We see one strong white-light flare around the 20-minute mark on night one, followed by smaller white-light flares on nights two and three. Unlike with the other objects, it also seems that there are flares in the Balmer light curve that are not represented in the TESS light curve. Note, for instance, the enhancements around the 35-minute and the 100-minute mark on night one, the 60-minute mark on night two, and the 80 and 100-minute mark on night three. This star was the only binary of the four observations.}
    \label{fig:BD220time}
\end{figure*}

Furthermore, the Balmer-line flares on the second night align with small (1-2\%) flares detected in the TESS light curve but have a significantly longer decay time. Lastly, these flares are also noted in the Balmer decrement, where we see decreases in the decrement during the flares. This decrease aligns with expectations from the literature \citep[e.g.][]{allred_radiative_2006}.





\subsubsection{TIC 220044948}

The plot Figure~\ref{fig:BD220time} follows the same format as Figure~\ref{fig:BD415time}. In Figure~\ref{fig:BD220time}, we plot the M2.5 dwarf \object{TIC 220044948}, which has a white dwarf companion on a 0.6-day orbit \citep[matching the photometric $P_\mathrm{rot}$;][]{baroch_carmenes_2021}. However, the companion is faint and its Balmer absorption is very broad due to pressure broadening. During night one, there was a flare in the Balmer lines around the 20-minute mark that matched the large, long-duration TESS flare.

\subsubsection{TIC 178947176}

\begin{figure}[htp!]
    \centering
    \includegraphics[width=\columnwidth]{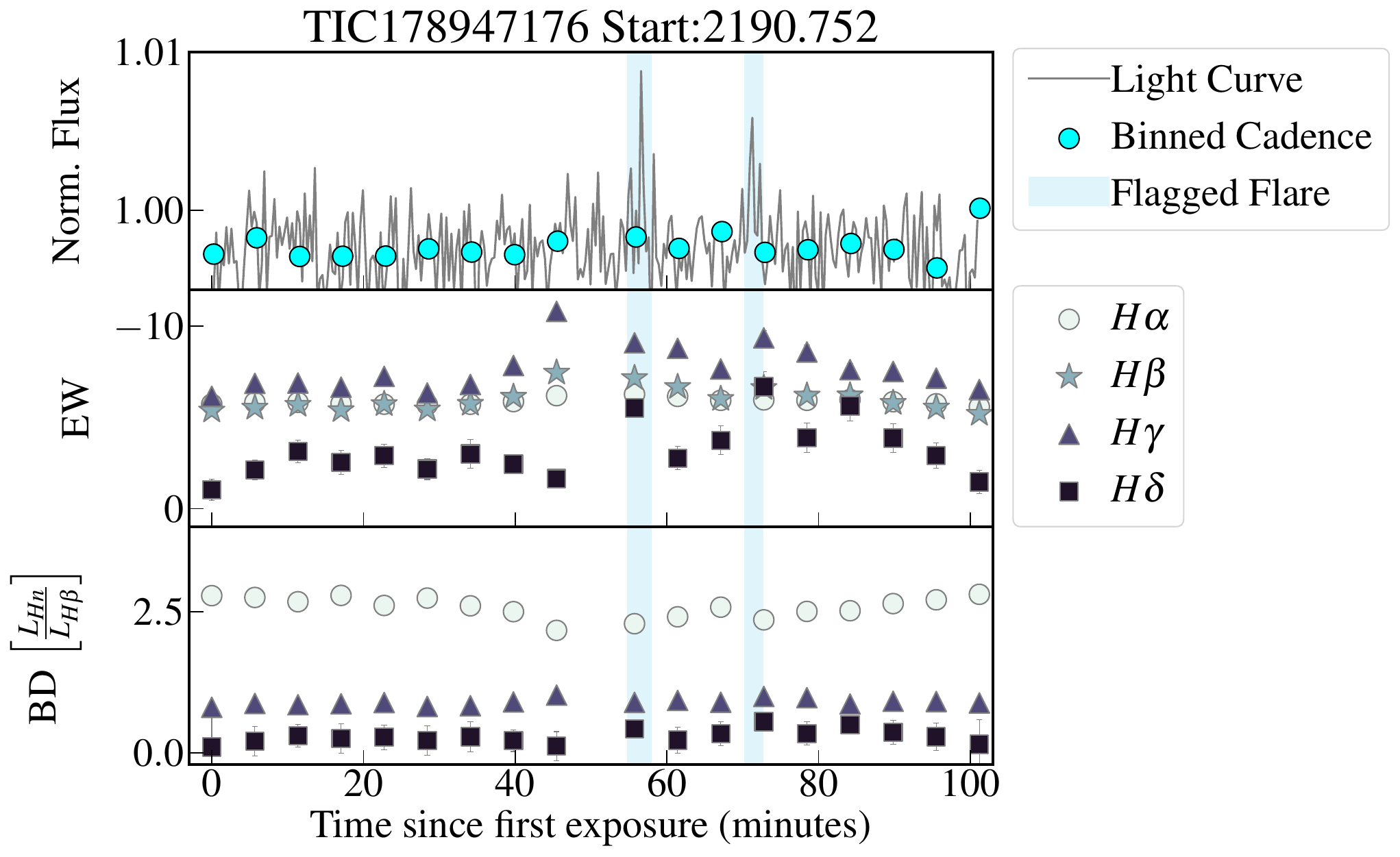}
    \includegraphics[width=\columnwidth]{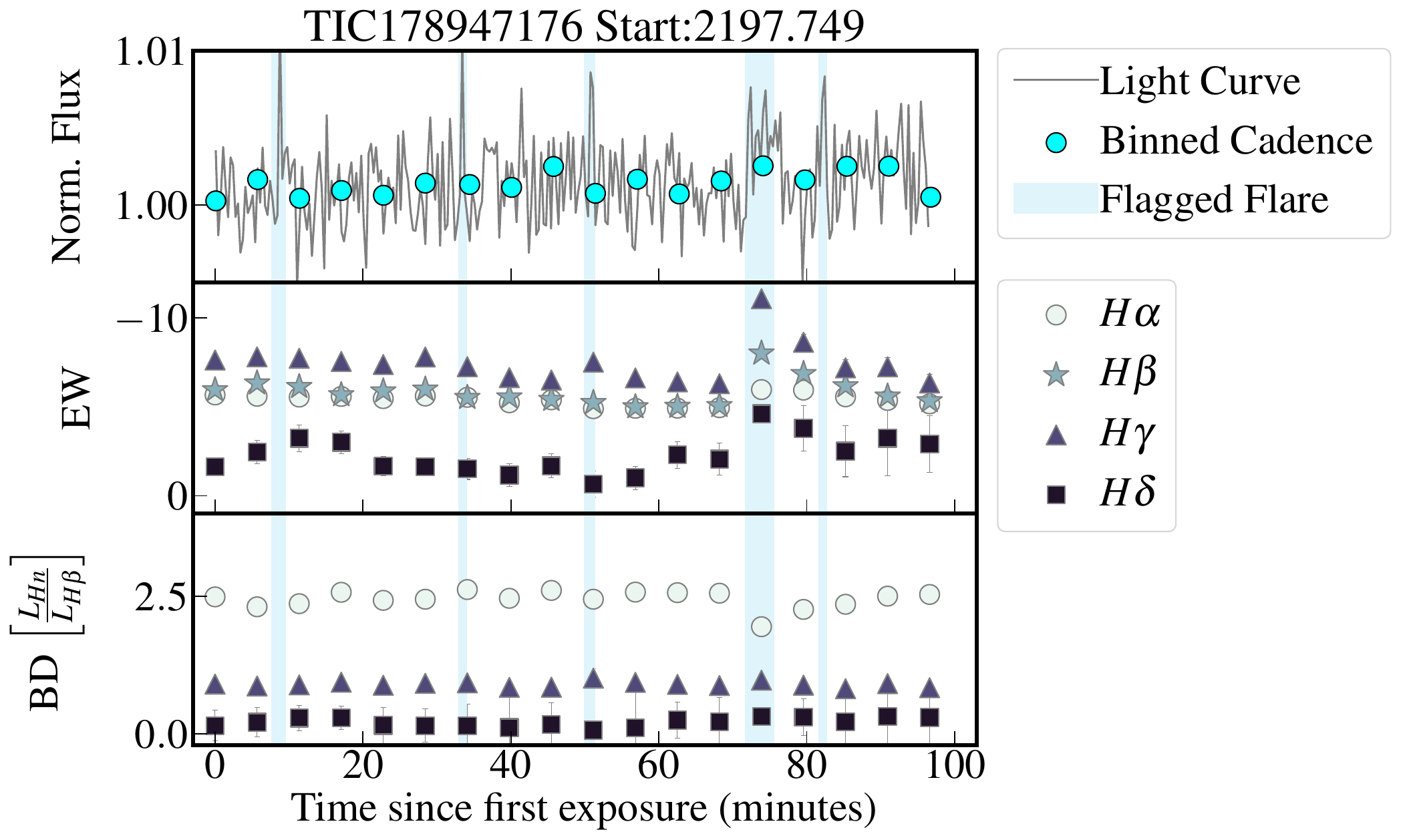}
    \caption{Same as Figure~\ref{fig:BD415time}, but for \object{TIC 178947176} (only two nights are shown). We see no clear enhancement from flares, except possibly in night two towards the 70-80 minute mark. This star has an observation length of 80. At the 50-minute mark, we removed data that was overwhelmed with systematics.}
    \label{fig:BD178time}
\end{figure}

In Figure~\ref{fig:BD178time}, we plot the data for the fully convective M Dwarf \object{TIC 178947176}. The cadence of the spectral observations is 300 seconds or 5 minutes and the TESS cadence is 20 seconds. The lower cadence of the spectroscopic data makes it more difficult to detect Balmer flares; however, we can see a possible enhancement from flares around the 80-minute mark on night two (day 2197) that corresponds to a TESS flare. 

\subsubsection{Time delays between TESS and Balmer line flares}\label{sec:timedelay}

In \object{TIC 220044948} and \object{TIC 415508270}, we see a short delay in the onset of the Balmer flare compared to the white-light flare (seen in the binned-cadence TESS data).
The delays are $<$5 min and we note that the offsets are present before we add the half exposure time to the OSMOS times. 

In the literature, there are examples of short-time delays between the onset flare in white light and the flares reflected in the spectra. For instance, the 2010 megaflare of the eruptive variable \object{YZ CMi} has a time delay between \hg \ and the continuum emission and a more gradual response on the successive flare events \citep{kowalski_white_2010,kowalski_time-resolved_2013}. Another example is a flare in the M dwarf \object{Proxima Cen} described in \citet{macgregor_discovery_2021}, 
which showed \ha \ peaking later than the TESS 120-second light curve.

Time delays in \ha \ line emission are also seen in the sun and are associated with slower thermal transport times \citep{canfield_h_1990}. Time delays of the \cak emission compared to both continuum and Balmer lines have also been observed and can be linked to the Neupert effect \citep{neupert_comparison_1968}. The Neupert effect describes the observed correlation between the total (nonthermal) emission produced during the impulsive phase of a flare and the total (thermal) emission produced during the gradual phase. In the scenario modeled by \citet{kowalski_time-resolved_2013},  back-warming of lower atmospheric layers results in a time delay between continuum emission and Ca II emission. This may also explain the delay between the Balmer lines, with bluer lines peaking later than redder lines during night one (Figure~\ref{fig:BD220time}, top left panel). 

\subsubsection{Differences between white-light and spectroscopic flares}\label{sec:sunflares} 

In Figures 6-8, there are flares in TESS or white-light flares with no corresponding Balmer lines flares--while \object{TIC 178947176} could be attributed to cadence, that is not the case for \object{TIC 415508270}. Moreover, we also see instances in which there was flaring activity in the Balmer lines with no corresponding white-light flare (see in particular \object{TIC 415508270} night two and \object{TIC 220044948} night one $\sim$ 30-minute mark and night two $\sim$ 60-minute mark). These could be attributed to non-white-light flares, flares seen in both the sun \citep{watanabe_characteristics_2017} and in other M dwarfs such as AD Leo \citep{namekata_optical_2020}. 

For AD Leo, \citet{namekata_optical_2020} observe two \ha \ flares without corresponding white-light flares\footnote{In these scenarios, they also noted that \hg \ and \hd \ exhibited larger EWs than \ha, as seen in our Figures 6–8.}. The authors model the stellar atmosphere using the \texttt{RADYN} package \citep{carlsson_non-lte_1992, carlsson_does_1995, carlsson_formation_1997, carlsson_dynamic_2002} (further discussed in \S\ref{sec:radyn_modeling}) and find that the relationship between \ha \ and the optical continuum is non-linear. This non-linearity likely arises from differences in optical thickness and emissivity between the two. Specifically, for decreasing chromospheric heating, white-light emission decreases by orders of magnitude more than \ha \ emission. Consequently, white-light emission becomes undetectable under low heating flux conditions. 

A similar result has also been seen in the Sun. Solar non-white-light flares are distinguished from white-light flares by a weaker coronal magnetic field strength \citep{watanabe_characteristics_2017}. White-light flares are also distinguished by shorter timescales and shorter flare ribbon distances. Our observations for \object{TIC 415508270} on night two around the 60-minute mark match the characteristics of a non-white-light flare. These are the strongest and longest-duration Balmer flares we observe but correspond to relatively weak and short-duration TESS flares.

\subsection{Spectroscopic rotation signatures in TIC 283866910}\label{sec:amber}

\begin{figure}[htp!]
    \centering
    \includegraphics[width=\columnwidth]{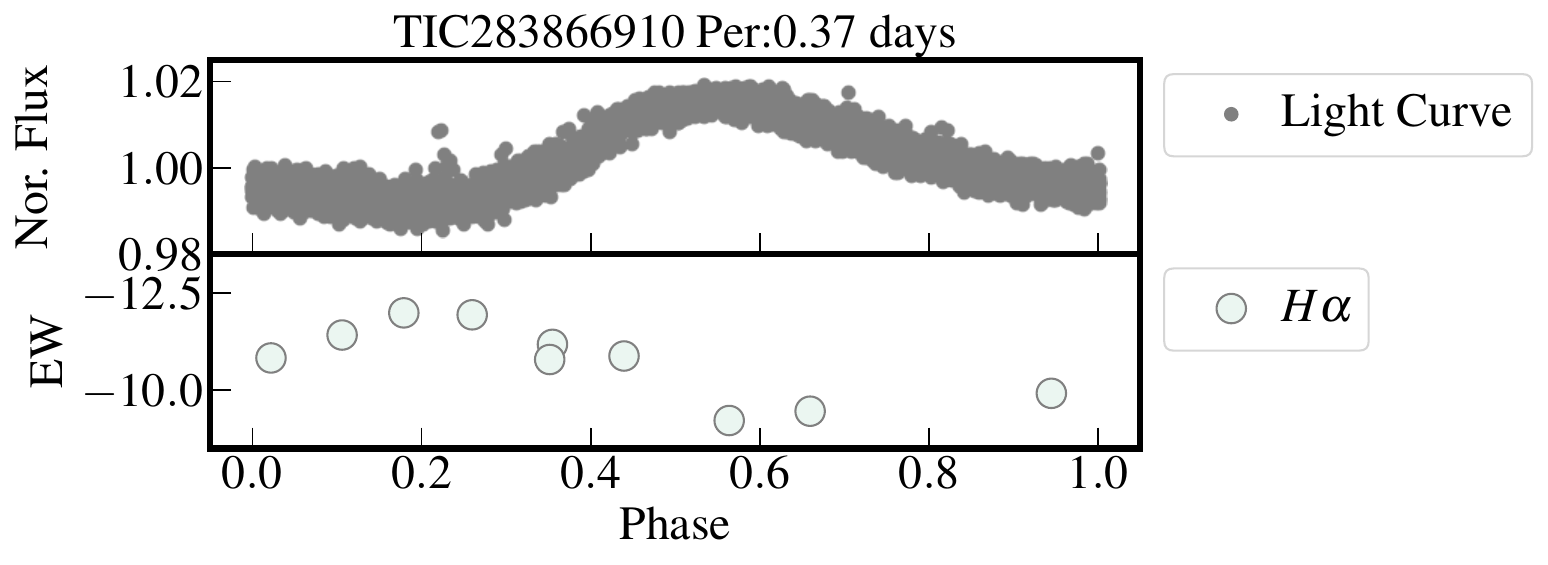}
    \includegraphics[width=\columnwidth]{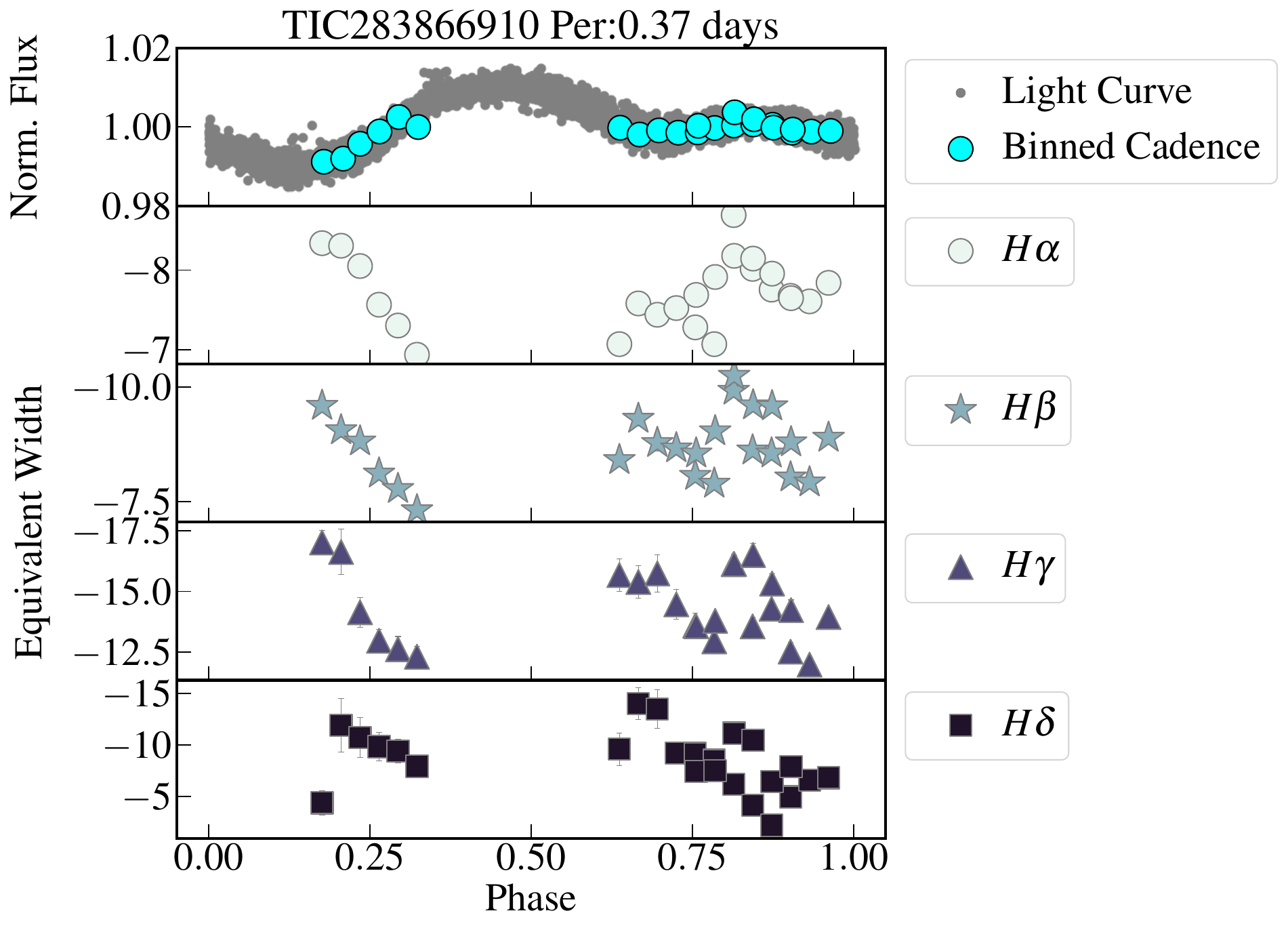}

    \caption{Top: TESS 120-second sector 5 light curve (semi-contemporaneous with \citet{medina_variability_2022} spectra; overlaps with the last two data points instead of all ten). We do not see a similar result to the findings in that paper.
    Bottom: TESS 120-second light curve, sector 32. Subplots 2-5: The equivalent width light curves for each detected Balmer line.  This object had a longer cadence than the rest, though we compared it to the $P_\mathrm{rot}$ instead of the flares. We do not get a full phase-fold but see the Balmer lines \ha \ and \hb \ following the anti-phase of the binned TESS light curve data.} 
\label{fig:phase283time}
\end{figure}

\object{TIC 283866910} is a fully convective M dwarf, of estimated radius: $0.271 \pm 0.008 R_\odot$, effective temperature: 3210 K, mass: $0.242 \pm 0.006 M_\odot$, and a rotation period of 0.37 days \citep{medina_variability_2022}.  It was also identified as a member of the AB Doradus young moving group and therefore has an age of about 150 Myr \citep{medina_variability_2022}.

Due to its faintness, \object{TIC 283866910}, the cadence for this star was 900 seconds or 15 minutes--too low to be able to catch any flare enhancements. Instead, we phase-folded its light curve (Figure~\ref{fig:phase283time}) to look for the rotation modulation. In their sample of 10 stars, \citet{medina_variability_2022} found that \object{TIC 283866910} was the only star in their sample to showcase a clear relationship between \ha \ and photometric modulation observed in MEarth photometry \citep{berta_transit_2012}. 
The origin of rotational variations observed in late M-dwarfs by TESS and Kepler--whether they stem from dark spots, bright spots, or a combination of both--remains an open question. For instance, the bright spot fraction for fully convective M-dwarfs, Trappist-1 \citep{radica_promise_2024} and GJ 1214 \citep{mallonn_gj_2018} are negligible while GJ 486 \citep{moran_high_2023} and LHS 1140 \citep{cadieux_transmission_2024} show a contribution of at least a 25\% contribution of faculae. The correlation or anti-correlation of photometric modulation and chromospheric emission could demonstrate whether active regions are bright or dark.

We replot the data from \citet{medina_variability_2022} in comparison to contemporaneous TESS data from Sector 5, and we plot our data in comparison to contemporaneous TESS data from Sector 32 in Figure~\ref{fig:phase283time}. All data are phase-folded to a period of 0.37 days using 2458423.7150, our assumption for the time zero-point used in that work. We note that we are unable to reproduce the figure in \citet{medina_variability_2022}, instead, we see larger \ha \ EW during the flux minimum in the light curve. This is consistent with the presence of a large active region that is darker than the photosphere.


\begin{deluxetable}{LLLLL}
\tablecaption{F-test to Compare \ha\ EW Variability and Rotational Phase for a Sub-sample of M Dwarfs}
\tablehead{\colhead{TIC} & \colhead{Month} & \colhead{F-test} & \colhead{p-value} & \colhead{CL \%}}
\startdata
283866910 & \text{2020 Dec} & 2.84 & 0.08 & 91.93 \\
220044948 & \text{2020 Dec} & 1.77 & 0.17 & 82.60 \\
415508270 & \text{2020 Dec} & 1.17 & 0.31 & 68.60  \\
178947176 & \text{2020 Dec} & 0.08 & 0.92 & 8.06 \\
266744225 & \text{2021 Jan} & 0.19 & 0.83  & 17.29 \\
65673065 & \text{2021 Jan} & 0.49 & 0.62 & 37.96 \\
275309165 & \text{2021 Feb-Mar} & 1.72 & 0.21 & 79.37\\
366323063 & \text{2021 Mar} & 0.36 & 0.70 & 29.98 \\
26126812 & \text{2021 Oct} & 3.04 & 0.07 & 92.66 \\
436880490 & \text{2021 Oct} & 1.15 & 0.33 & 66.80\\
90768237 & \text{2021 Oct} & 0.93 & 0.40 & 59.54 \\
405461329 & \text{2021 Oct} & 2.60 & 0.11 &  89.05\\
435308532 & \text{2021 Oct} & 1.62 & 0.23 & 76.87
\enddata
\label{tab:ftest}
\tablecomments{Following the F-test described in \S~4 of \citet{medina_variability_2022}, we calculate the F-test value, the p-value, and the confidence level (CL) given the $P_\mathrm{rot}$ of each star is listed in Table~\ref{tab:bigtab}.}
\end{deluxetable}

As in \citet{medina_variability_2022}, we perform an F-test (Equation 4 therein) to help determine if there is a correlation between the $P_\mathrm{rot}$ and the \ha\ EW. The null hypothesis is that there is no correlation, and the equivalent width is consistent described by Equation 2 in \citet{medina_galactic_2022}. The alternative hypothesis is that there is a correlation, described by Equation 3 \citet{medina_variability_2022}. For \object{TIC 283866910}, these authors find a significant correlation with rotation period--F-test statistic value of 19.98 and a 98.3\% confidence. For the data shown in Figure~\ref{fig:phase283time}, we find F-test = 33 at a p-value of 0.0002 (significant at 99.97\% confidence). For this object using our data from 2020 Dec, we find F-test = 3 at a p-value of 0.08 (marginally significant at 91.93 \% confidence). In Table~\ref{tab:ftest}, we show all of the F-test statistics for \ha\ for the stars with more than six spectra. 

\section{Radiative hydrodynamic flare modeling with RADYN} \label{sec:radyn_modeling}

The intensity and shape of the Balmer line profiles provide constraints on the electron beam heating environment that sets the relative emission at ultraviolet to infrared wavelengths. However, forward modeling to derive the properties of the electron beam from flare spectroscopy has only recently become possible with the development of a new grid of self-consistent 1D radiative hydrodynamic (RHD) flare models that incorporate non-local thermodynamic equilibrium (non-LTE) conditions for a broad range of injected beam intensities and energies into the RADYN code \citep{kowalski_time-dependent_2024}. We therefore forward model the observed \ha \ and \hb \ spectra, for \object{TIC 415508270} night two, using a grid search spanning all 43 RADYN model runs with electron fluxes of 10$^{10}$ to 10$^{13}$ erg s$^{-1}$ cm$^{-2}$.

The \texttt{RADYN} code \citep{carlsson_does_1995, carlsson_formation_1997} self-consistently solves the non-LTE radiative transport and hydrodynamics equations for the stellar atmosphere during plasma heating by a non-thermal electron beam \citep{allred_radiative_2006, allred_unified_2015}. Each beam consists of a power-law distribution of energetic electrons described by its maximum electron flux in erg s$^{-1}$ cm$^{-2}$, minimum electron energy in keV, and the index of accelerated electrons $\delta$ that sets the relative abundance of high and low-energy electrons. Each model follows a naming convention such as mF11-85-3 where the electron flux is 10$^{11}$ erg s$^{-1}$ cm$^{-2}$, the cutoff energy is 85 keV, and the power law index $\delta$ is 3. The initial conditions of the stellar atmosphere at the top of the corona are set by a flare loop half-length of 10$^9$ cm, $T_\mathrm{eff} \approx$ 3600 K, the gas temperature of 5 MK, log $g$ = 4.75, the ambient electron density of 3$\times$10$^{10}$ cm$^{-3}$, and a uniform cross-sectional area. The atmospheric response and flare spectrum are computed at 0.2 s time resolution during a pulsed injection consisting of a 1 s increase and 10 s decrease in beam intensity as described in \citet{aschwanden_pulsed_2004}. At each time step, the flare continuum is sampled at 95 points from 6--40,000 \AA, and hydrogen line emission is computed at 2 \AA \ resolution with the pressure broadening profiles of \citet{tremblay_spectroscopic_2009}. Finally, a fiducial flare spectrum is obtained for each model run by averaging across the full 10 s duration.

\begin{figure*}[htp!]
	\centering
        \subfigure
	{
		\includegraphics[width=0.85\textwidth]{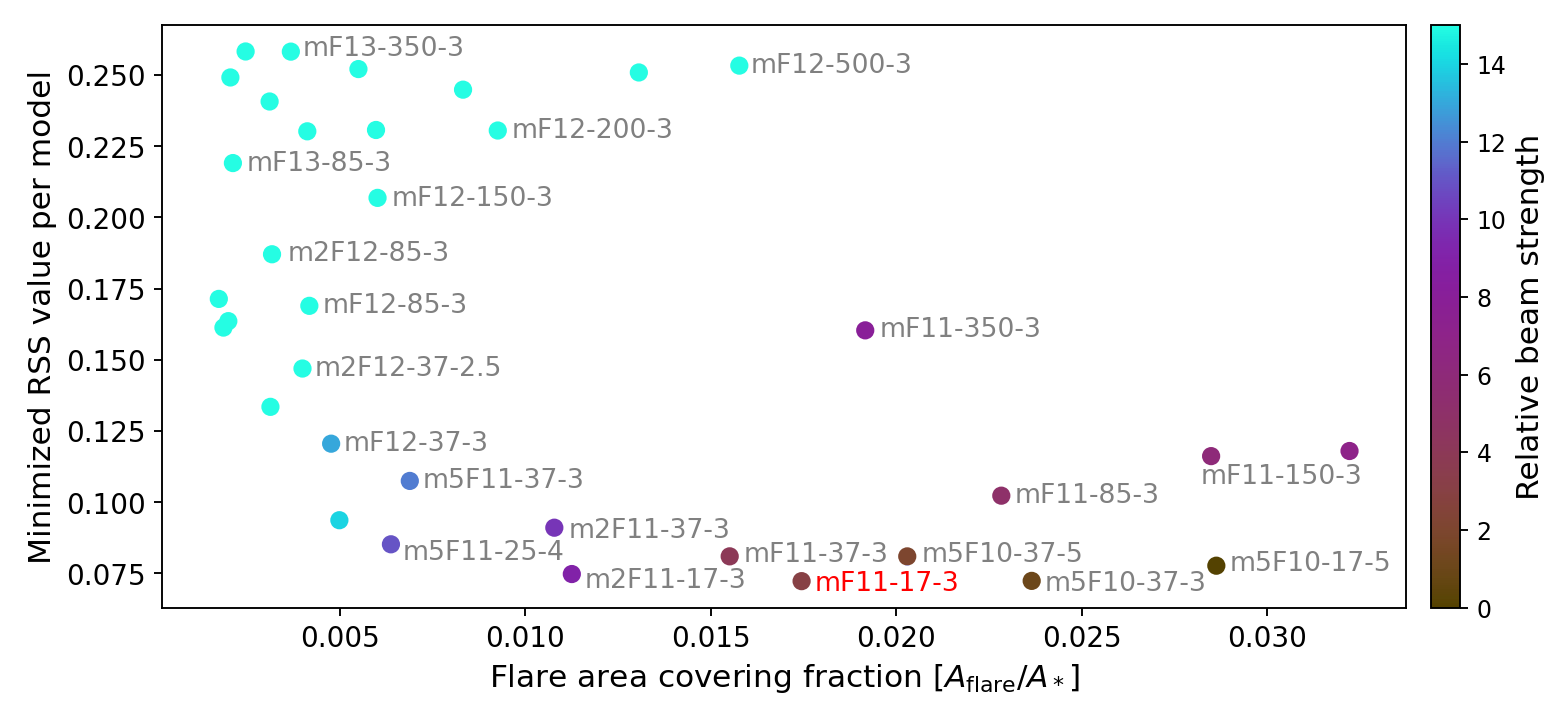}
	}
        \subfigure
	{
		\includegraphics[width=0.85\textwidth]{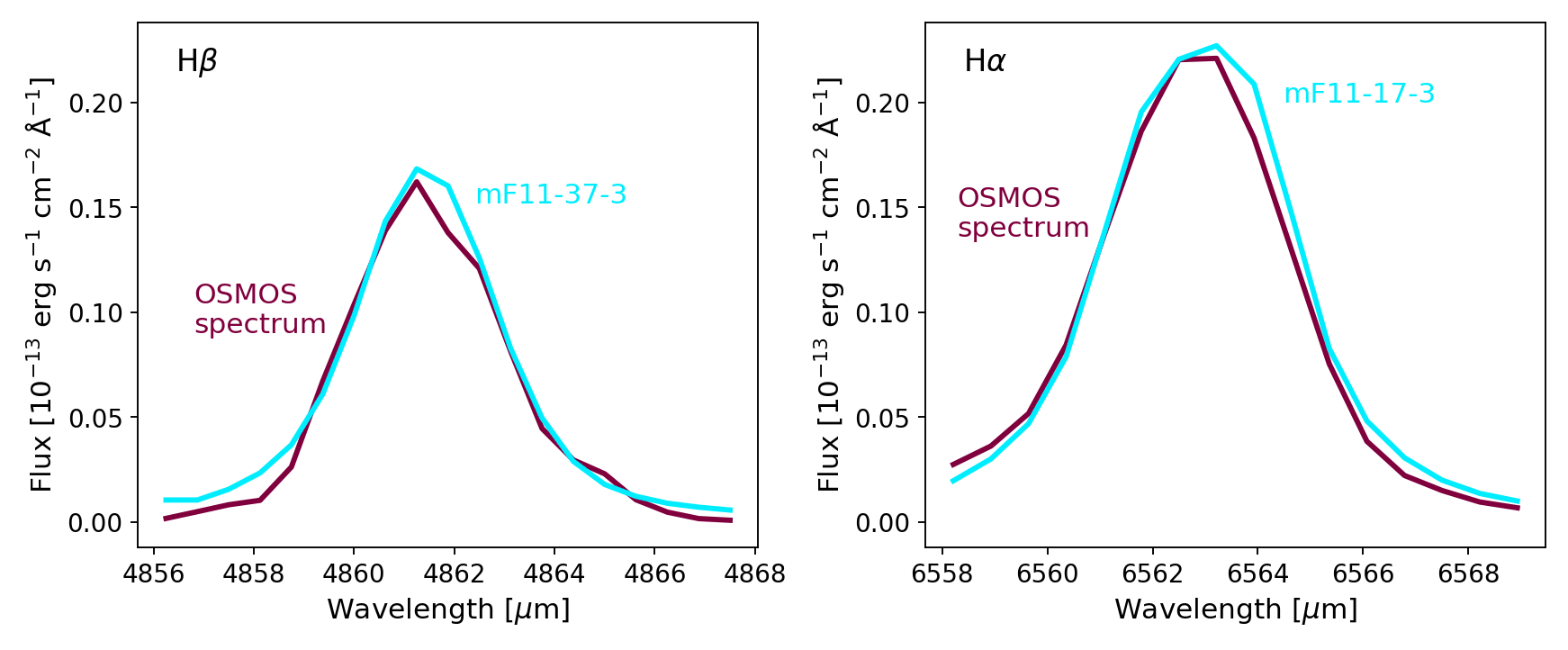}
	}
	\caption{Top: Grid search results comparing 43 RADYN model fits to the \ha\ and \hb\ lines during the peak of the largest flare in our sample, which was observed with OSMOS from \object{TIC 415508270} on 2020-12-09 at 5:52 UT. The electron beams that give the lowest RSS fits the observed spectra are characterized by low electron fluxes and energies of 1--2$\times$10$^{11}$ erg s$^{-1}$ cm$^2$ and 17--37 keV, respectively. The flare area coverage versus RSS value points are color-coded by increasing beam intensity, with select models labeled in gray. The model labeled in red signifies the favored models based on our data. Bottom: The \ha \ and \hb \ spectral profiles favor the mF11-17-3 and m(2)F11-37-3 RADYN models in the single-line fits, respectively. The \ha \ line is the dominant contributor to the joint \ha \ and \hb \ fit selection.}
	\label{fig:radyn_model}
\end{figure*}

We use Equation 3 of \citet{howard_characterizing_2023} to fit the observed spectral flux density of the line emission using two RADYN spectral components representing the compact flare kernel and extended flare ribbon, where the relative contributions of the two model components are determined by the areal covering fractions of each component. For each fit, we first flux-calibrate the observed spectra against an M5V PHOENIX model from the MUSCLES archive version 23 \citep{wilson_mega-muscles_2021}, computed from the Lyon BT-Settl CIFIST 2011\_2015 grid \citet{allard_phoenix_2016}. At the same time, a 1D Gaussian blur is applied to the RADYN line profile outputs for the \ha \ and \hb \ lines to smooth the double-peaked line profile of the RADYN models to match the single peak observed in the data, which we interpret to be a consequence of the lower effective resolution of the observations. The maximum-likelihood fit parameters for each RADYN model are identified using 10,000 Monte Carlo trials to measure the distribution of the residual sum of squares (RSS) values as a function of flare area with a uniform prior of 0--0.1 on the flare covering fraction for each component. The 0.1 upper limit on the covering fraction is imposed based on the observed TESS band flare energy of $E_\mathrm{TESS}$ = 7.4$\times$10$^{30}$ erg, implying an active region covering fraction less than 0.04 \citep{howard_evryflare_2019}. The first set of MC trials for each model always converged on a covering fraction of zero for one of the model components, so we reran the MC trials using only one model component and free parameter. The maximum-likelihood covering fraction and corresponding minimum RSS value are recorded for separate MC fits for the \ha \ line, \hb \ line, and joint fits to both lines. Uncertainties in the covering fraction are computed with 200 bootstrap measurements of the minimum RSS value randomly sampled from 50\% of the MC distribution.

Our grid search for the \ha-only fit favors the mF11-17-3 model with a covering fraction of 0.036 $\pm$ 0.002 and the m2F11-17-3 model with a covering fraction of 0.024 $\pm$ 0.004 (both with RSS = 0.0015). The H$\beta$-only grid search favors the m2F11-37-3 model with a covering fraction of 0.009 $\pm$ 0.006 (RSS = 0.0007) and mF11-37-3 model with a covering fraction of 0.012 $\pm$ 0.006 with RSS values of 0.0007--0.0008. The joint \ha \ and \hb \ grid search favors the mF11-17-3 model with a covering fraction of 0.017 $\pm$ 0.005 (RSS = 0.0722) and m5F10-37-3 model with a covering fraction of 0.024 $\pm$ 0.003 (RSS = 0.0723). The similarity in the lowest RSS electron beam models compared to the multiple order of magnitude range of tested electron flux densities and cutoff energies supports the interpretation of the flare emission due to a 10$^{11}$ erg s$^{-1}$ cm$^{-2}$ electron beam consisting primarily of 10--40 keV electrons. The mF11-17-3 model predicts a peak luminosity in the TESS band of 1.4 $\pm$ 0.1$\times$10$^{28}$ erg s$^{-1}$ that is similar to the TESS band luminosity of 1.7 $\pm$ 0.5$\times$10$^{28}$ erg s$^{-1}$ observed during the H$\alpha$ and H$\beta$ peak time, lending additional support for this interpretation. We note an mF11-17-3 beam and 0.036 $\pm$ 0.002 covering fraction produces a peak flare luminosity of 1.4 $\pm$ 0.1$\times$10$^{27}$ erg s$^{-1}$ at FUV and 1.9 $\pm$ 0.1$\times$10$^{28}$ erg s$^{-1}$ at NUV wavelengths, suggesting simultaneous TESS, \ha, and \hb \ observations may be useful in constraining the photochemical radiation environment during M Dwarf flares.

\section{Discussion and Summary}\label{sec:conclude}

In this paper, we present a sample of 93 targets (Table~\ref{tab:bigtab}) that spectroscopic data using OSMOS and/or ModSpec and photometry from TESS. Although we only analyze the 77 M dwarfs in this work, we publish all of the observed data, including several T Tauri stars, as supplemental material. 

Using the TESS data, we derive $P_\mathrm{rot}$ values using Gaussian Processes for stars without a period in the literature and calculate the rotation modulation amplitudes using \rvar. We use the spectra to measure Balmer line luminosities using new $\chi$-factor values--the ratio of the flux of continuum near the spectral line to bolometric luminosity--for \ha\ following the technique from \citet{nunez_factory_2024}; differences between the $\chi$ values in that paper and this one are due to the difference in continua regions. In comparison to the $\chi$ values in \citet{walkowicz_characterizing_2008}, our values of $\chi_\mathrm{H\alpha}$ and $\chi_\mathrm{H\beta}$ are similar for cooler temperatures until $T_\mathrm{eff} \approx 3100$ K. 

Our analysis of the 77 M dwarfs is summarized below.
\begin{itemize}
    \item In \ref{sec:amp}, we update the amplitude-activity plot from \citetalias{garcia_soto_contemporaneous_2023}, now including \hb, \hg, and \hd. The analysis for \ha\ includes 35 likely single stars, 33 for \hb, 26 for \hg, and 20 for \hd. For \hg\ and \hd, we exclude stars with visible systematic issues (e.g., cosmic rays or fringing). We find weak, positive correlations but the p-values do not indicate that these are significant. Both sample size and intrinsic variability affect the relation.
    \item In \S~\ref{sec:invarb}, we find that higher-order Balmer lines exhibit larger intrinsic variability, consistent with previous work \citep{duvvuri_fumes_2023}. This analysis includes 61 stars (with more than 6 spectra) and excludes those with visible systematic issues.
    \item In \S\ref{sec:flares}, we observe flares in photometry and spectroscopy for 3 stars. There is no strong concordance between flares observed in TESS and the Balmer lines. The differences in these morphologies could reflect variations in the heating and cooling rates of each flare, with one potentially exhibiting similar rates while the other shows different behavior. The Balmer decrement becomes shallower during flares detected in the Balmer lines. 
    \item In \S\ref{sec:timedelay}, we observe evidence of a short flare onset delay in the Balmer flares and potential pre-dimming before the flare shown in Figure~\ref{fig:BD415time}. The time delay may be related to the Neupert effect.
    \item In \S\ref{sec:amber} we look for evidence of rotation signals in \ha. For 13 stars with TESS data, we conduct an F-test, and all stars \ha \ EW are not correlated with the photometry over a rotation period. However, we note that observations do not usually span a full $P_\mathrm{rot}$. Our re-analysis of data of the young M dwarf \object{TIC 283866910} from \citet{medina_variability_2022} and analysis of our new data show that flux in the TESS bandpass is significantly anti-correlated with \ha \ luminosity. This suggests the presence of an active region that is darker than the photosphere.
    \item In \S\ref{sec:radyn_modeling} we use the RADYN code to model the effects of electron beams on the flare emission noticed in \ha\ and \hb\ for \object{TIC 415508270}. We perform a grid search across a range of model parameters and use Monte Carlo simulations to determine the best-fitting model, revealing that the flare is likely caused by an electron beam with a flux of flux around $\approx$10$^{11}$ erg s$^{-1}$ cm$^{-2}$ and electron energy in the range 10--40 keV. Simultaneous multi-wavelength data would further help constrain the photochemical radiation environment.
\end{itemize}

\vspace{2mm}
\noindent We thank all the people who have made the paper possible including Adam Kowalski, Amber Medina, Arcelia Hermorsillo Ruiz, Becky Flores, Bur\c{c}in Mutlu-Pakdil, Graham Edwards, John Thorstensen, Kevin Covey, Sebastian Pineda, Zachory Berta-Thompson, and Yuta Notsu. We thank Tony Negrete and Eric Galayda and proxy observer Justin Rupert at the MDM Observatory in Arizona. 

This work has made use of data from the European Space Agency (ESA) mission
{\it Gaia} (\url{https://www.cosmos.esa.int/gaia}), processed by the {\it Gaia}
Data Processing and Analysis Consortium {\it Gaia } (DPAC, 
\url{https://www.cosmos.esa.int/web/gaia/dpac/consortium}). Funding for the DPAC
has been provided by national institutions, in particular, the institutions
participating in the {\it Gaia} Multilateral Agreement. This paper also includes data collected by the {\it TESS} mission. TESS was obtained from the Mikulski Archive for Space Telescopes (MAST) at the Space Telescope Science Institute. The data analyzed can be accessed via \dataset[https://doi.org/10.17909/t9-st5g-3177]{https://doi.org/10.17909/t9-st5g-3177} and \dataset[https://doi.org/10.17909/t9-nmc8-f686]{https://doi.org/10.17909/t9-nmc8-f686}. STScI is operated by the Association of Universities for Research in Astronomy, Inc., under NASA contract NAS5–26555. Support to MAST for these data is provided by the NASA Office of Space Science via grant NAG5–7584 and by other grants and contracts. Funding for the {\it TESS} mission is provided by the NASA's Science Mission Directorate.

\facilities{Hiltner(OSMOS), McGraw-Hill(ModSpec), TESS}

\software{\astropy \ \citep{astropy_collaboration_astropy_2013,astropy_collaboration_astropy_2018,astropy_collaboration_astropy_2022}, 
\celerite \  \citep{foreman-mackey_celerite_2017,dan_foreman-mackey_dfmcelerite_2020},
\texttt{edmcmc} \ \citep{vanderburg_avanderburgedmcmc_2021},
\ernlib \ (\url{https://github.com/ernewton/ernlib}),  
\texttt{exoplanet} \citep{dan_foreman-mackey_exoplanet-devexoplanet_2020,agol_analytic_2020},
\IRAF \ \citep{tody_iraf_1986,tody_iraf_1993,national_optical_astronomy_observatories_iraf_1999},
\texttt{Llamaradas-Estelares} (\url{https://github.com/lupitatovar/Llamaradas-Estelares/tree/main}),
\lightkurve \ \citep{lightkurve_collaboration_lightkurve_2018, geert_barentsen_keplergolightkurve_2020},  
\texttt{M\_-M\_K-} \citep[\url{https://github.com/awmann/M_-M_K-};][]{mann_how_2019},
\texttt{NumPy} \citep{oliphant_guide_2006}, 
\pymc \ \citep{salvatier_probabilistic_2016},
PyRAF \citep{science_software_branch_at_stsci_pyraf_2012},
\readmultispec \ \citep{kevin_gullikson_general-scripts_v10_2014}, 
\texttt{SciPy} \citep{virtanen_scipy_2020}, 
\texttt{specutils} \citep{nicholas_earl_astropyspecutils_2019}, 
\starspot \ \citep{angus_ruthangusstarspot_2021}, 
\stella \ \citep{feinstein_stella_2020}, 
\texttt{thorosmos} (\url{https://github.com/jrthorstensen/thorosmos}), 
\texttt{thorsky} (\url{https://github.com/jrthorstensen/thorsky})}

\bibliography{references}{}
\bibliographystyle{aasjournal}

\end{document}